\documentclass[twocolumn]{aastex61}

\usepackage{multirow}
\usepackage{amsmath}

\def\sygma{\texttt{SYGMA}}
\def\omegac{\texttt{OMEGA}}
\def\omegap{\texttt{OMEGA+}}

\newcommand{\iso}[2]{\hbox{${}^{#1}{\rm #2}$}}

\shorttitle{GCE of Radioactive Isotopes}
\shortauthors{C\^ot\'e et al.}

\begin{document}

\title{Galactic Chemical Evolution of Radioactive Isotopes}

\correspondingauthor{Benoit C\^ot\'e}
\email{benoit.cote@csfk.mta.hu}

\author[0000-0002-9986-8816]{Benoit C\^ot\'e}
\affiliation{Konkoly Observatory, Research Centre for Astronomy and Earth Sciences, Hungarian Academy of Sciences, Konkoly Thege Miklos ut 15-17, H-1121 Budapest, Hungary}
\affiliation{Joint Institute for Nuclear Astrophysics - Center for the Evolution of the Elements (JINA-CEE)}
\affiliation{NuGrid Collaboration, \url{http://nugridstars.org}}

\author{Maria Lugaro}
\affiliation{Konkoly Observatory, Research Centre for Astronomy and Earth Sciences, Hungarian Academy of Sciences, Konkoly Thege Miklos ut 15-17, H-1121 Budapest, Hungary}
\affiliation{Monash Centre for Astrophysics, School of Physics and Astronomy, Monash University, VIC 3800, Australia}

\author[0000-0002-3855-5816]{Rene Reifarth}
\affiliation{Goethe University Frankfurt, Max-von-Laue-Str. 1, Frankfurt, 60438, Germany}
\affiliation{NuGrid Collaboration, \url{http://nugridstars.org}}

\author[0000-0002-9048-6010]{Marco Pignatari}
\affiliation{E.A. Milne Centre for Astrophysics, University of Hull, Hull, HU6 7RX}
\affiliation{Konkoly Observatory, Research Centre for Astronomy and Earth Sciences, Hungarian Academy of Sciences, Konkoly Thege Miklos ut 15-17, H-1121 Budapest, Hungary}
\affiliation{Joint Institute for Nuclear Astrophysics - Center for the Evolution of the Elements (JINA-CEE)}
\affiliation{NuGrid Collaboration, \url{http://nugridstars.org}}

\author{Blanka Vil\'agos}
\affiliation{Konkoly Observatory, Research Centre for Astronomy and Earth Sciences, Hungarian Academy of Sciences, Konkoly Thege Miklos ut 15-17, H-1121 Budapest, Hungary}

\author{Andr\'es Yag\"ue}
\affiliation{Konkoly Observatory, Research Centre for Astronomy and Earth Sciences, Hungarian Academy of Sciences, Konkoly Thege Miklos ut 15-17, H-1121 Budapest, Hungary}

\author[0000-0003-4446-3130]{Brad K. Gibson}
\affiliation{E.A. Milne Centre for Astrophysics, University of Hull, Hull, HU6 7RX}
\affiliation{Joint Institute for Nuclear Astrophysics - Center for the Evolution of the Elements (JINA-CEE)}

\begin{abstract} 

The presence of short-lived ($\sim$\,Myr) radioactive isotopes in meteoritic inclusions at the time of their formation represents a unique opportunity to study the circumstances that led to the formation of the Solar System. To interpret these observations we need to calculate the evolution of radioactive-to-stable isotopic ratios in the Galaxy. We present an extension of the open-source galactic chemical evolution codes NuPyCEE and JINAPyCEE that enables to track the decay of radioactive isotopes in the interstellar medium. We show how the evolution of isotopic ratio depends on the star formation history and efficiency, star-to-gas mass ratio, and galactic outflows. Given the uncertainties in the observations used to calibrate our model, our predictions for isotopic ratios at the time of formation of the Sun are uncertain by a factor of 3.6. At that time, to recover the actual radioactive-to-stable isotopic ratios predicted by our model, one can multiply the steady-state solution (see Equation~1) by $2.3^{+3.4}_{-0.7}$. However, in the cases where the radioactive isotope has a half-life longer than $\sim$\,200\,Myr, or the target radioactive or stable isotopes have mass- and/or metallicity-depended production rates, or they originate from different sources with different delay-time distributions, or the reference isotope is radioactive, our codes should be used for more accurate solutions. Our preliminary calculations confirm the dichotomy between radioactive nuclei in the early Solar System with $r$- and $s$-process origin, and that $^{55}$Mn and $^{60}$Fe can be explained by galactic chemical evolution, while $^{26}$Al cannot.

\end{abstract}




\section{Introduction} \label{sec:intro}

Radioactive isotopes with half-lives longer than $\sim$0.1\,Myr offer a wide range of opportunities for investigating stellar and explosive nucleosynthesis, galactic evolution and mixing in the interstellar medium (ISM), and the conditions existing at the time of birth of the Sun \citep{radiobook2018}. Many of these long-lived isotopes are well known to have been present in the early Solar System, via analysis of meteoritic rocks and inclusions. Depending on which time intervals their half-lives are comparable to, they can be used to measure the age of events of cosmological, astrophysical, and planetary interest. For example, the age of our Galaxy \citep{dauphas05} and of the Sun \citep{amelin02,2017GeCoA.201..345C} can be measured using radioactive isotopes whose half-lives are on the order of Gyrs, such as those of U and Th. Radionuclides with half-lives of the order of tens of Myr, such as \iso{182}Hf and \iso{129}I, can be used to measure the time of formation and chemical differentiation of asteroids and planets \citep{kleine02}. They can also probe the time when the molecular cloud in which the Sun formed became isolated from galactic nucleosynthetic additions, the so-called ``isolation time'' \citep[e.g.,][]{wasserburg06,huss09,lugaro14,lugaro16,vescovi18}.

The Galactic abundances of two of the most short-lived radioactive isotopes of interest here, \iso{26}Al (0.72 Myr) and \iso{60}Fe (2.62 Myr), are observed via $\gamma$-ray spectroscopy and reflect the signature of fresh nucleosynthetic events in the Milky Way \citep{diehl13}. Their abundances in the early Solar System can be used as tracers of the environment where the Sun was born \citep[see review by][]{LUGARO2018} and the heat generated by the radioactive decay of \iso{26}Al affected the thermo-mechanical evolution of planetesimals \citep{lichtenberg16b}. The energy generated by the decay of the U and Th isotopes, and by \iso{40}K, is also responsible for a significant fraction of the heat budget of the Earth's interior, and possibly of extra-solar terrestrial rocky planets \citep{unterborn15}. 

The modelling of the evolution of radioactive isotopes relative to stable isotopes in the Galaxy is a main ingredient required for interpreting these observations and exploiting their implications. 
Here we present open-source galactic chemical evolution (GCE) codes dedicated to the evolution of radioactive isotopes, which can be freely employed to study any abundance ratios of interest. We analyze quantitatively many of the dominant uncertainties in GCE that can affect the results, and make a direct comparison with the traditional analytical GCE model of \citet{clayton88}.

In the following Subsection \ref{sec:previous} we describe previous work done in the context of GCE of radionuclides. In Section~\ref{sec:equations} we introduce our codes, in Section~\ref{sec:results} we present the resulting radioactive-to-stable abundance ratios and analyze the impact of uncertainties in the star formation history, the gas-to-star mass fraction, galactic outflows, and delay times. We present a GCE best fit model as well as a range of possible solutions and how these compare to the results obtained using the traditional steady-state approach. In Section~\ref{sec:disc} we discuss the uncertainties and the limitations of our framework and present two examples of its application: the ratio of the short-lived \iso{26}Al and \iso{60}Fe and that of the very long-lived \iso{235}U and \iso{238}U. In Section~\ref{sec:conc} we present a summary, conclusion, and future work.

\subsection{Previous work}\label{sec:previous}

Radioactive and stable isotopes are produced together in the Galaxy by stars, supernovae, and events emerging from binary interactions, the only difference being that radioactive nuclei decay with time. Typically, it can be considered that radioactive nuclei reach a steady-state abundance in the ISM, provided by the balance between their stellar production rate and their decay rate. Simply, the more abundant the radioactive nucleus is, the more it decays until there is no variation in its abundance. The exact value of the half-life affects this evolution as the shorter the half-life the quicker the steady-state abundance is reached. It is more interesting, however, for comparison to observations to calculate abundance ratios. 
In this case, we need to investigate the galactic evolution of a radioactive isotope relative to another radioactive isotope, or to a stable isotope. The abundance of a stable isotope after a certain galactic time $T_{\rm Gal}$ can be considered to be simply given by its stellar production rate multiplied by the time considered. From these considerations it can be derived that

\begin{equation}
\label{eq:steady}
\frac{N_{\rm radio}}{N_{\rm stable}}=
\frac{P_{\rm radio}}{P_{\rm stable}} \, \frac{\tau}{T_{\rm Gal}},
\end{equation}
where $N_{\rm radio}$ and $N_{\rm stable}$ are the abundances of the radioactive and the stable nuclei, respectively, at time $T_{\rm Gal}$, $P_{\rm radio}$ and $P_{\rm stable}$ their constant stellar production rates, and $\tau$ the mean life of the radioactive isotope, related to the half-life $t_{1/2}$ by $t_{1/2}=\tau\,\mathrm{ln}(2)$.
This simple formula is based on steady-state and constant stellar production rates. This approximation, that does not have to be true for all radioactive isotopes, allows one to derive radioactive-to-stable isotopic ratios at any given time in the Galaxy, such as at the time of the formation of the Sun. It has been traditionally and extensively employed to derive the isolation time \citep[see, e.g.,][]{wasserburg06,huss09}. However, as already pointed out by \citet{clayton85,clayton88}, and further developed by \cite{huss09}, there are several complications that need to be taken into account.

First, the Galaxy is well known to not be a ``closed box'', meaning that inflow of primordial or low-metallicity gas is required to explain its features (e.g., \citealt{1980FCPh....5..287T}), in particular the stellar metallicity distribution function. \citet{clayton85} already included this effect in his analytical description of GCE, which results in the introduction of a multiplication factor ($k+1$) in Equation~(\ref{eq:steady}), where $k$ is a free parameter that sets the temporal profile of the infall rate.  This multiplication factor accounts for the fact that the infall modifies the star formation rate and that radioactive and stable isotopes are more affected by the local and the integrated star formation rate, respectively. It also accounts for the fraction of the abundances of stable isotope locked inside old stars. The value of the infall parameter $k$ has been found to be in the range $1-3$ in order to match observational constraints \citep{clayton84,clayton88}. A more recent attempt at deriving the value of $k$ based on astronomical observations resulted in $2.7\,\pm\,0.4$ \citep{dauphas03}. 

Second, stellar production rates are not constant but can change with metallicity. This was considered in detail by \citet{huss09}, who developed an analytical description of this effect within the framework of the analytical GCE Clayton models. This resulted in different multiplications factors to the steady-state solution described in Equation~(\ref{eq:steady}) of ($k+1$), ($k+2$), or ($k+1$)($k+2$) depending if the radioactive and the stable isotopes are both primary, both secondary, or one of each type, respectively. 

While the introduction of infall, the use of astronomical constraints to determine the related free parameters, and the improved treatment of the metallicity-dependence of the stellar production rates are clear improvements from the simple steady-state, closed-box formula of Equation~(\ref{eq:steady}), the description of the Galaxy in all these previous work has still been performed analytically. One limitation of the analytical approach is that not all possible infall prescriptions can be solved analytically. Another limitation is that one value for the stellar production rate has to be used together with the different multiplication factors, while stellar yields may behave in more complex ways than a simple primary or secondary trend. These effects can be fully captured using numerical GCE models, which provide more accurate results than analytical models. GCE models can deal with any type of infall prescriptions, and because they can use metallicity- and mass-dependent stellar yields, they offer a stronger connection with nuclear astrophysics and stellar nucleosynthesis. In addition, these models can keep track of all the different sources that could simultaneously contribute to the target isotopes and account for the fact that the nucleosynthetic contribution from some stellar sources is subjected to certain delay times. Finally, the possibility of now observed galactic outflows has still not been considered yet in relation to the evolution of radioactive-to-stable isotopic ratios in the Galaxy. 

Only a few studies have addressed the evolution of radioactive isotopes in a fully numerical GCE context. \citet{timmes95} considered the evolution of \iso{26}Al and \iso{60}Fe, using mass-dependent core-collapse supernova yields, to estimate their current injection rate and total mass in the ISM. Using metallicity-dependent yields, \citet{travaglio14} considered four isotopes produced exclusively by the $p$ process (\iso{92}Nb, \iso{97,98}Tc, and \iso{146}Sm) along with their stable reference isotopes, also produced by the $p$ process, under the assumption that Chandrasekhar-mass Type~Ia supernovae are the only producer on these isotopes in the Galaxy.  \citet{sahijpal14} considered five radioactive isotopes (\iso{26}Al, \iso{36}Cl, \iso{41}Ca, \iso{53}Mn, and \iso{60}Fe) with very short half-lives between 0.1 and 3.7\,Myr, using mass- and metallicity-dependent yields. 
However, none of these studies quantified the effect of GCE uncertainties on the evolution of the radioactive-to-stable ratios. Also, of the codes used for these studies, to our knowledge only that by \citet{timmes95} is publicly available. Our aim is to make substantial progress on the GCE of radioactive isotopes by providing open-source codes and a detailed analysis of the effect of GCE uncertainties.

\section{Chemical Evolution Codes} 
\label{sec:equations}
The treatment of radioactive isotopes has been implemented in the open-source JINA-NuGrid chemical evolution pipeline \citep{2017nuco.confb0203C}. This numerical framework is based on object-oriented programming such that each code (or module) available within the pipeline can be used independently or be introduced into more complex systems.  In the next sections, we briefly review the chemical evolution codes and describe how the radioactive isotope implementation has been jointed to the framework. All codes are publicly available and are part of the NuPyCEE\footnote{\url{http://github.com/NuGrid/NuPyCEE}} and JINAPyCEE\footnote{\url{http://github.com/becot85/JINAPyCEE}} packages on GitHub. Documentation on how to use the codes is provided in the form of iPython Jupyter notebooks and is cited in the following subsections. Although installing the code is relatively straightforward, the installation can be bypassed by using the online virtual \texttt{cyberhubs}\footnote{\url{http://wendi.nugridstars.org}} environment \citep{2018ApJS..236....2H}.

Although they do not include a treatment for radioactive isotopes, we refer to \citet[\texttt{flexCE}]{2017ApJ...835..224A} and \citet[\texttt{Chempy}]{2017A&A...605A..59R} for alternative open-source chemical evolution codes.

\subsection{Simple Stellar Population Model}
\label{sec:sygma}
The \sygma\ code (\citealt{2017arXiv171109172R}, Stellar Yields for Galactic Modeling Applications) calculates the mass of isotopes ejected by an entire population of stars as a function of time (see also \citealt{1999ApJS..123....3L,2009MNRAS.399..574W,2017AJ....153...85S}). All stars are assumed to form at the same time from the same parent cloud of gas and to inherit the same initial chemical composition. \sygma\ includes the contribution of massive stars, low- and intermediate-mass stars, Type~Ia supernovae, neutron star mergers, as well as an arbitrary number of additional enrichment sources that can be defined by the user\footnote{\url{https://github.com/NuGrid/NuPyCEE/blob/master/DOC/Capabilities/Delayed_extra_sources.ipynb}}. Each individual source is weighted by an initial mass function and has its own nucleosynthetic yields that can be mass- and metallicity-dependent. The code accounts for the lifetime (or delay-time distribution) of every enrichment sources independently.

\subsection{Galaxy Model with Inflows and Outflows}
\label{sec:omega}
The \omegac\ code (\citealt{2017ApJ...835..128C}, One-zone Model for the Evolution of GAlaxies)  calculates the evolution of the chemical composition of the gas inside a galaxy.  From a given star formation history (SFH), the code creates several stellar populations throughout the lifetime of the galaxy and follows the combined contribution of all stars on the enrichment process.  Each stellar population has its own properties and is modeled using \sygma\ (Section~\ref{sec:sygma}).  As in all one-zone models, \omegac\ adopts the homogeneous-mixing approximation.  This means that once the stellar ejecta is deposited in the galactic gas, it is instantaneously and uniformly mixed within the gas reservoir. We refer to \cite{2003PASA...20..401G}, \cite{2008EAS....32..311P}, \cite{2012ceg..book.....M}, and \cite{2013ARA&A..51..457N} for more details on the basics of GCE simulations.

\omegac\ includes galactic inflows and outflows in order to consider, in a simplified way, the interactions between galaxies and their surrounding environment (see e.g., \citealt{2015ARA&A..53...51S,2017MNRAS.470.4698A,2017ARA&A..55...59N}). Inflows introduce gas into the galaxy, fuel star formation, and usually dilute the gas metallicity inside the galaxy (e.g., \citealt{2017ASSL..430..221F}).  Outflows on the other hand expel gas from the galaxy (e.g., \citealt{2005ARA&A..43..769V,2016ApJ...819...29B,2018MNRAS.473.4077P}).  For galaxies with masses similar or lower than the Milky Way, those outflows are mainly driven by stellar feedback (e.g., \citealt{2012MNRAS.421.3522H,2015ARA&A..53...51S,2018arXiv181100558Z}).

Within our framework, the evolution of the total mass of gas ($M_\mathrm{gas}$) inside a galaxy is described as \citep{1980FCPh....5..287T,1997nceg.book.....P,2012ceg..book.....M}
\begin{equation}
\dot{M}_\mathrm{gas}(t)=\dot{M}_\mathrm{inflow}(t)+\dot{M}_\mathrm{ej}(t)-\dot{M}_\star(t)-\dot{M}_\mathrm{outflow}(t),
\end{equation}
where the four rate terms on the right-hand side represent the mass added by galactic inflows, added by stellar ejecta, locked away by star formation, and lost by galactic outflows, respectively. In addition to the total mass of gas, the code keeps track of individual isotopes. The total number of isotopes included in the calculation is only limited by the number of isotopes available in the input stellar yields.

Some representative inflow prescriptions are explored in Section~\ref{sec:SFH}, but more options are available within our framework\footnote{\url{https://github.com/becot85/JINAPyCEE/blob/master/DOC/OMEGA\%2B_defining_gas_inflow.ipynb}}. The stellar ejecta is calculated by summing the contribution of every stellar populations formed by time $t$,

\begin{equation}
    \dot{M}_\mathrm{ej}(t)=\sum_j\dot{M}_\mathrm{ej}^j(M_j, Z_j, t-t_j),
\end{equation}
where $\dot{M}_\mathrm{ej}^j$ is the mass ejected by the $j$th stellar population, and $M_j$, $Z_j$, and $t_j$ are the initial mass, initial metallicity, and formation time of that population. The $t-t_j$ quantity refers to the age of the $j$th population at time $t$. One population of stars is created per timestep in the simulation, and their initial mass and metallicity is set by the star formation rate (SFR) and chemical composition of the galactic gas at that time.

The SFR in our model is directly proportional to the mass of gas inside the galaxy, and is defined by (e.g., \citealt{2001MNRAS.328..726S,2006RPPh...69.3101B})
\begin{equation}
    \label{eq:sfe}
    \dot{M}_\star(t)=\frac{\epsilon_\star}{\tau_\star}M_\mathrm{gas}(t)=f_\star M_\mathrm{gas}(t),
\end{equation}
where $\epsilon_\star$ and $\tau_\star$ are the dimensionless star formation efficiency and star formation timescale, respectively. In this work, we combine these two quantities into $f_\star$, the star formation efficiency in units of yr$^{-1}$. Here we assume that $f_\star$ is constant with time, but we refer to \cite{2017arXiv171006442C} for alternative prescriptions. The outflow rate is assumed to be proportional to the star formation rate, and defined as (e.g., \citealt{2005ApJ...618..569M,2015MNRAS.454.2691M})
\begin{equation}
    \label{eq:eta}
    \dot{M}_\mathrm{outflow}(t)=\eta\dot{M}_\star(t),
\end{equation}
where $\eta$ is the mass-loading factor regulating the strength of the outflow. In this work, we assume that $\eta$ is constant with time, but more options are available within our framework\footnote{\url{https://github.com/becot85/JINAPyCEE/blob/master/DOC/OMEGA\%2B_defining_gas_outflow_galactic.ipynb}}.

\subsection{Circumgalactic Medium and Recycling}
\label{omega_p}
The \omegap\ code (\citealt{2017arXiv171006442C}) is a two-zone model and represents a simple extension of \omegac\ that allows to follow the chemical evolution of the circumgalactic medium (CGM) as well as the chemical evolution inside the galaxy. In practical terms, \omegap\ consists of a large gas reservoir surrounding an \omegac\ object, the latter representing the galaxy. Using \omegap\ instead of \omegac\ allows to keep track of the isotopes ejected by galactic outflows, and to reintroduce  them at later times into the galaxy via galactic inflows (see, e.g., \citealt{2008MNRAS.387..577O,2017MNRAS.470.4698A,2018arXiv180807872C}). The evolution of the total mass of gas ($M_\mathrm{CGM}$) in the CGM is described as
\begin{equation}
  \begin{aligned}
\dot{M}_{\rm CGM}(t)& =  \dot{M}_{\rm CGM,in}(t)\,+ \\ & \dot{M}_{\rm outflow}(t) - \dot{M}_\mathrm{inflow}(t) - \dot{M}_{\rm CGM,out}(t),
\label{eq_main_cgm}
  \end{aligned}
\end{equation}
where the four rate terms on the right-hand side represent the mass accreted from the intergalactic medium into the CGM, added by galactic outflows, lost by galactic inflows, and expelled from the CGM into the intergalactic medium. The latter medium represents the space outside the volume occupied by the CGM, which is typically defined by a sphere with a radius equals to the virial radius of the dark matter halo hosting the central galaxy. In this work, we ignore the interaction between the CGM and the intergalactic medium and set $\dot{M}_{\rm CGM,in}$ and $\dot{M}_{\rm CGM,out}$ to zero at all times. We refer to our online documentation\footnote{\url{https://github.com/becot85/JINAPyCEE/blob/master/DOC/OMEGA\%2B_list_of_parameters.ipynb}} and to \cite{2017arXiv171006442C} for details on how to activate such interaction.

\subsection{Decay of Radioactive Isotopes}
\label{decay_module}
The new version of our codes allows to use both stable and radioactive yields for any enrichment source. When including radioactive yields, the gas reservoir of the galaxy is split into a stable and radioactive components, which are then followed separately. Once isotopes are present in the radioactive gas component, each one of them is decayed following their specific decay properties. If the decay products are stable isotopes, they are transferred into the stable gas component. The decay occurs during the chemical evolution calculations, which means that radioactive isotopes are continuously added by stellar ejecta.

Our framework offers two options for dealing with the decay of radioactive isotopes, which are described in the next subsections.  Details on how to activate and use those options with our codes are given in our online documentation\footnote{\url{https://github.com/NuGrid/NuPyCEE/blob/master/DOC/Capabilities/Including_radioactive_isotopes.ipynb}}.

\subsubsection{Single Decay Channel Using an Input File}
The simplest option is to provide an input file that lists all the radioactive isotopes that will be included in the calculation. There is no limit on the number of isotopes that can be included. For each one of them, the half-life and the isotope in which the specie decays into must be provided. This option can only be used when the target radioactive isotopes have a single decay channel, meaning that their decay product only consists of one isotope (e.g., $^{26}$Al~$\rightarrow$~$^{26}$Mg). In this case, the decay of a radioactive isotope $i$ is calculated by

\begin{equation}
    \dot{N_i}(t) = - \frac{N_i(t)}{\tau_i},
\end{equation}
where $N_i$ and $\tau_i$ represent the abundance of isotope $i$ in number and its mean life, respectively.

\subsubsection{Multiple Decay Channels Using the Decay Module}
The second option is to use our decay module, an independent code originally programmed in Fortran that is now imported into our GCE codes. This module allows to decay isotopes with a single decay channel like $^{26}$Al as well as the ones that have multiple decay channels like $^{40}$K and $^{238}$U. In the module, the decay rates and channels are assumed to be the same as under terrestrial conditions, where many experimental data exist. The reaction rates and branching ratios in the network are taken from the NUDAT Nuclear data files provided by the National Nuclear Data Center (\citealt{NUD07}). The network solver currently includes 22 decay channels: 

\begin {itemize}
	\item $\beta^{-}$, $\beta^{+}$/EC (the latter stands for electron capture),
	\item spontaneous emission of neutrons, protons, or alpha particles,
	\item spontaneous emission of 2 neutrons, 2 protons, or 2 alpha particles,
	\item $\beta^{-}$-delayed 1-, 2-, 3-, 4-neutron, neutron-alpha emission 
	\item $\beta^{+}$/EC-delayed 1-, 2-proton, proton-alpha emission
	\item $\beta^{-}$- and $\beta^{+}$/EC-delayed alpha emission,
	\item $\beta^{-}$-delayed 2-alpha emission,
	\item internal transition (de-excitation of isomers),
	\item $^{12}$C emission,
	\item spontaneous fission.
\end{itemize}

The module uses a publicly available Fortran subroutine of the GEF code \cite[GEneral description of Fission observables][]{2016NDS...131..107S,SJS17} to estimate the mass distribution of spontaneous fission events after the scission point. We used the approximation described in \cite{VHZ01} to determine the mass differences of neighboring isotopes, which is required in the treatment of the de-excitation and neutron emission of the fragments after scission.

An important aspect of the code is the correct treatment of decay chains. Long-lived isotopes like $^{238}$U (half-life of 4.47\,Gyr) decay on the same timescale as the galactic evolution. The decay products, however, can have much shorter half-lives. For example, in the following decay chain,

\begin{equation}
    ^{238}\mathrm{U}\,(\alpha)\,^{234}\mathrm{Th}\,(\beta^-)\,^{234}\mathrm{Pa}\,(\beta^-)\,^{234}\mathrm{U},
\end{equation}
$^{234}$Th has a half-life of 24~days, and $^{234}$Pa has a half-life of 6.7~hours for the ground state and 1.2~minutes for the isomer. For astrophysical applications, the accurate prediction of the equilibrium abundance is an important aspect. Indeed, since the decay activity of the corresponding isotopes can sometimes be observed, the abundance of the long-lived mother (here $^{238}$U) can be determined. An example is the observation of the decay of $^{60}$Co (half-life of 5.3~days) in the Milky Way and the derived abundance of its long-lived mother $^{60}$Fe \citep{HKJ05}. The approximate abundance ratio between the long-lived mother and the short-lived daughter (here $^{60}$Co) in equilibrium is

\begin{equation}\label{eq:equilibrium}
    \frac{N_\mathrm{long}}{N_\mathrm{short}} \approx \frac{\lambda_\mathrm{short}}{\lambda_\mathrm{long}}
\end{equation}
where $N$ is the abundance of a given (short- or long-lived) radioactive isotope and $\lambda=1/\tau$ is its decay constant.

For each radioactive isotope, the decay module solves the following equation,

\begin{equation}
    \dot{N}(t) = P - \lambda N(t),
    \label{eq:decay}
\end{equation}
where $P$ is the production rate coming from the decay of parent isotopes. Addition of isotopes by stellar ejecta in the ISM is treated in the GCE codes separately, not in the decay module. The stellar ejecta production term is therefore not included in  $P$. If the half-live of the isotope is much longer than the integration timestep $\Delta t$, Equation~(\ref{eq:decay}) can be solved step-wise. For $\lambda\Delta t<10^{-3}$, we assume a constant production and decay rates during the timestep, and the change in abundance can be expressed as

\begin{equation}
    \Delta N = P \Delta t - \lambda N \Delta t.
\end{equation}

For the daughters of long-lived isotopes, we solved the linear equation explicitly assuming a constant production rate,

\begin{equation}
    N(t+\Delta t) = \frac{P}{\lambda} \left(1-\mathrm{e}^{-\lambda \Delta t} \right) + N(t) \mathrm{e}^{-\lambda \Delta t}.
\end{equation}
The number of decays is derived by integrating $\lambda N(t)$ during the timestep, and the change of abundance becomes
\begin{equation}
    \Delta N = \left(\frac{P}{\lambda}-N\right) \left(1-\mathrm{e}^{-\lambda \Delta t} \right).
\end{equation}
This approach results in the equilibrium solution $N(t)\approx P/\lambda$ even if the time steps are much longer than the half-life time. This solution corresponds to Equation~(\ref{eq:equilibrium}). 

\begin{figure}
\center
\includegraphics[width=3.35in]{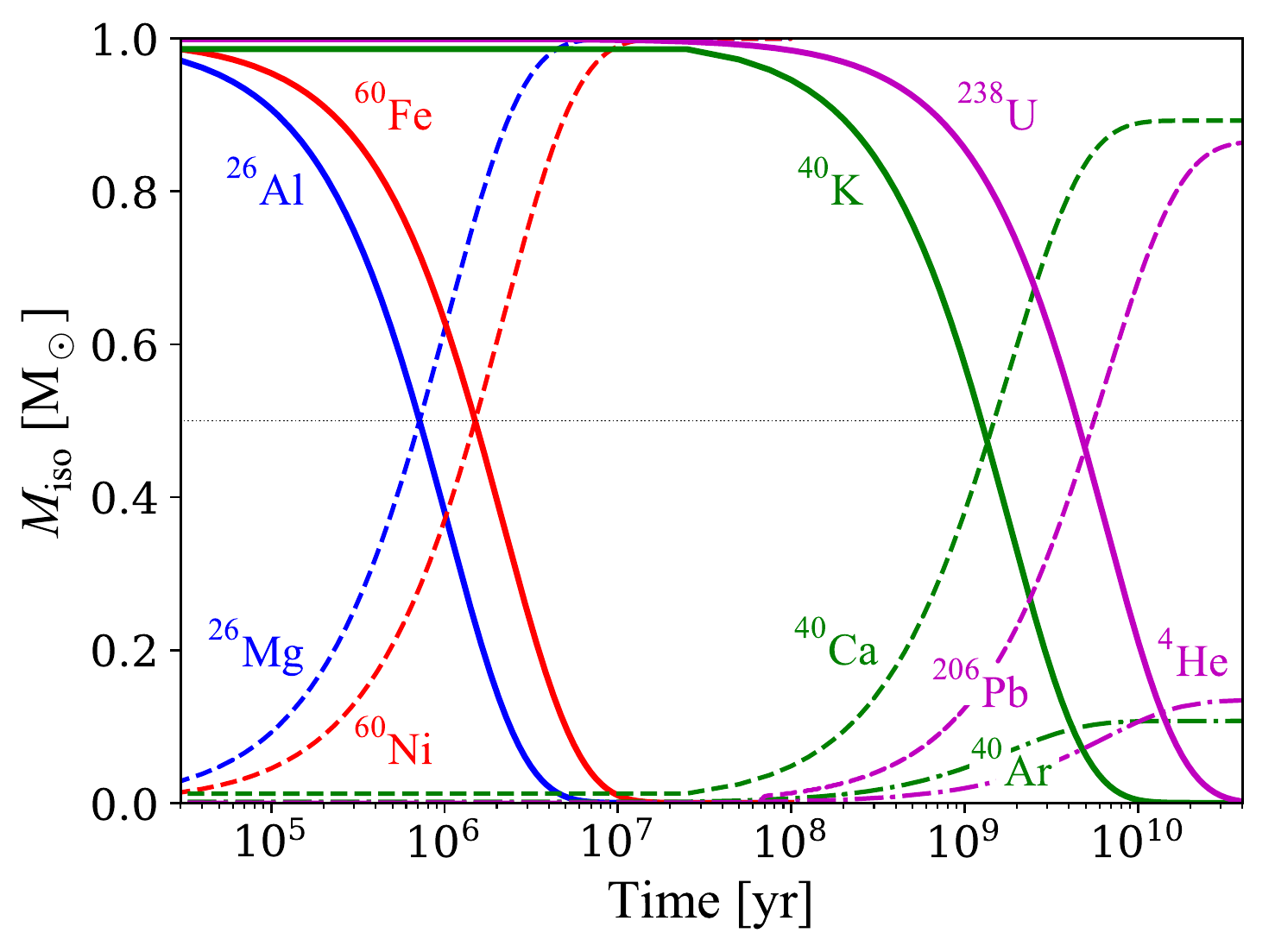}
\caption{Evolution of the mass of radioactive isotopes (solid lines) along with the rise of their daughter isotopes (dashed and dot-dashed lines), using the decay module included in our chemical evolution codes (Section~\ref{decay_module}). Each color represent a decay process of a specific radionuclide, as indicated on the figure.  There is no enrichment process in this figure, and all radioactive isotopes have initially been set to 1\,M$_\odot$. The time at which a solid line crosses the thin grey horizontal line represents the half-life of the associated radioactive isotope. 
\label{fig:decay}}
\end{figure}

As an example, Figure~\ref{fig:decay} shows the free decay of 1\,M$_\odot$ of $^{26}$Al, $^{60}$Fe, $^{40}$K, and $^{238}$U, calculated by the decay module implemented in our GCE framework. Some isotopes like $^{26}$Al decay into a single isotope, while others like $^{40}$K decay into two isotopes. $^{238}$U has a very complicated decay process that includes, among other channels, the emission of alpha particles ($^{4}$He). After 1\,Gyr of free decay, $^{238}$U has produced 566 different stable and radioactive isotopes, the most abundant being $^{206}$Pb and $^{4}$He. We remind that during a galaxy evolution calculation, the decay module is called at each timestep to decay the content of the radioactive gas component, which is continuously replenished by stellar ejecta. The mass of isotopes are converted into number back and forth at each GCE timestep to allow communication between the GCE codes and the decay module.

\section{Evolution of Isotopic Ratios in the Galaxy}
\label{sec:results}
As mentioned in Section~\ref{sec:intro}, the abundance of a radioactive isotope is usually measured relative to a reference stable isotope. Here we calculate the evolution of an isotopic ratio $M_\mathrm{radio}/M_\mathrm{stable}$, where $M$ is the mass of the respective isotopes in the ISM of the Galaxy. The goal is to explore how this evolution is affected by the input assumptions made in our GCE model \omegap. This will be used to quantify the confidence level of our predictions, given the uncertainties in the observations used to calibrate our Milky Way model. Our model targets the Galactic disk, not the halo or the bulge.


\subsection{Definition of the Numerical Experiment}
\label{sec:num_setup}
In this paper, the radioactive and stable isotopes under consideration and the astronomical event producing them are arbitrary.  Our goal is to provide a general understanding of the impact of galaxy evolution assumptions on the evolution of the  $M_\mathrm{radio}/M_\mathrm{stable}$ ratio. Although the exact value of the isotopic ratio does depend on the adopted nucleosynthetic yields and on the half-life of the radioactive isotope, the range of the predictions, i.e., the level of uncertainty, is insensitive to these quantities, as long as the half-life is significantly shorter than the lifetime of the Galaxy ($\sim$\,13\,Gyr).  Therefore, our results can be applied to any isotopic ratio and to any astronomical event (e.g., core-collapse supernovae, compact binary merger, asymptotic giant branch star, etc.).  As a reference, our results have been calculated with a half-life of 10\,Myr and a radioactive-to-stable mass ratio of 0.2 for the yields.

In the next sections, we compare our results to the analytic model of \cite{clayton84,clayton88}, hereafter referred to as Clayton's model, since this model has been widely used in the cosmochemistry community to calculate the chemical abundances of short-lived radioactive isotopes at the time of the formation of the Sun (e.g., \citealt{2000SSRv...92..133M,dauphas03,huss09,2013RPPh...76f6201R}).  In Section~\ref{sec:SFH}, to provide a consistent comparison, we initially simplify our chemical evolution model to mimic the conditions adopted in Clayton's model.  This includes a fixed gas-to-star mass fraction for a given SFH, no galactic outflow, and no delay between the formation of the progenitor stars and the ejection of the yields.  In Sections~\ref{sec:sfe}, \ref{sec:outflows}, and \ref{sec:DTD}, we relax those limitations one by one.  In Section~\ref{sec:uncertainty}, we present our best Milky-Way model along with its uncertainties, and compare our results with the steady-state formula.

Throughout this paper, $t_\odot$ refers to the formation time of the Sun in our Galactic disk simulation. The Universe is 13.8\,Gyr old \citep{2018arXiv180706209P} but galaxies only started to form a couple of 100\,Myr after the Big Bang \citep{2011ARA&A..49..373B}. The exact formation time of the Galactic disk is not precisely known. As a first order approximation, we thus ran all of our models for 13\,Gyr. Knowing the Sun formed 4.6\,Gyr ago \citep{2017GeCoA.201..345C}, we set $t_\odot\sim8.5$\,Gyr. 

\subsection{Shape of the Star Formation History}
\label{sec:SFH}

\begin{figure}
\center
\includegraphics[width=3.35in]{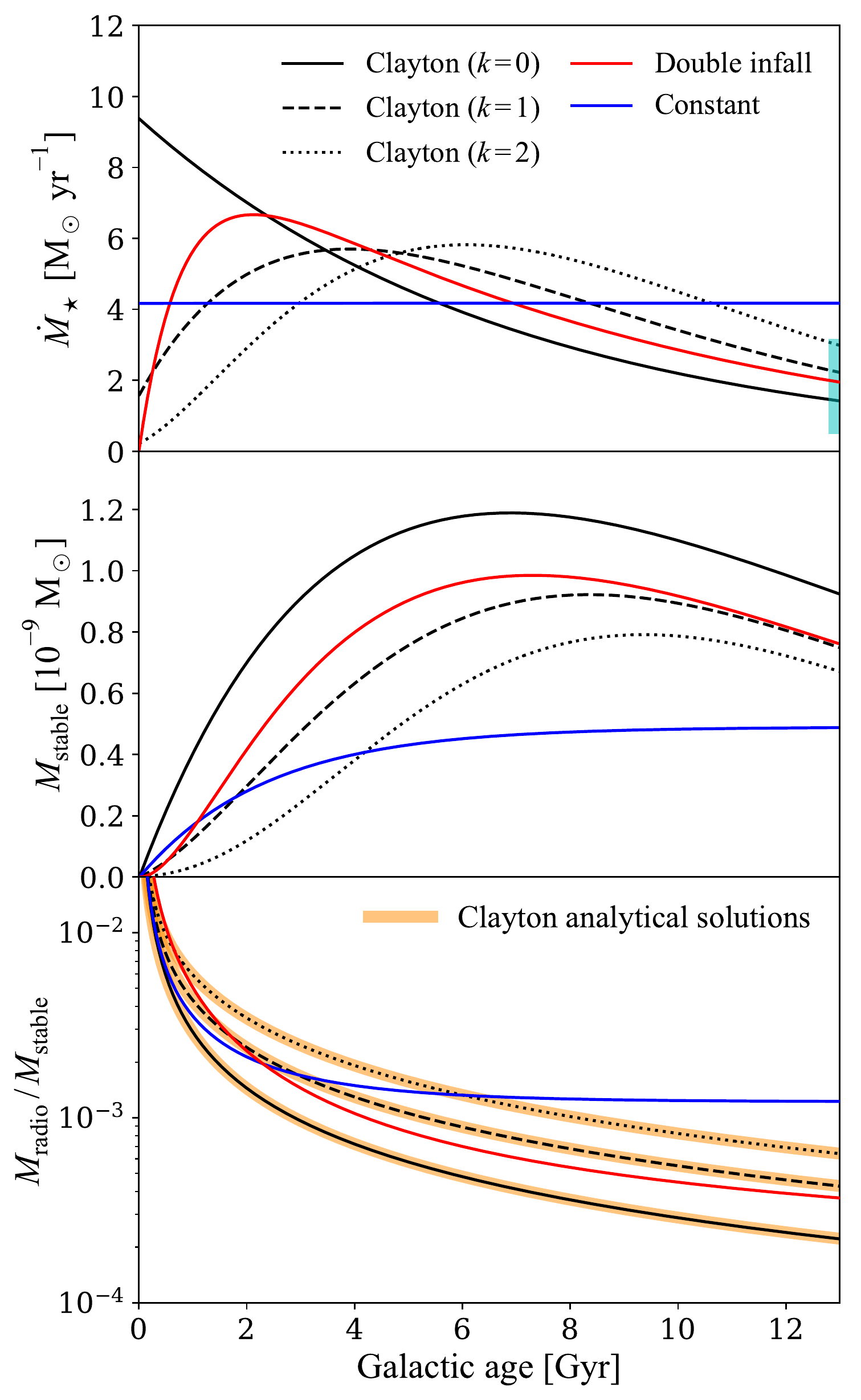}
\caption{Top panel: Star formation histories calculated assuming different galactic inflow histories: Clayton's model (black lines), the two-infall prescription of \citet[red line]{1997ApJ...477..765C}, and a constant inflow history (blue line). The vertical thick cyan line represents the current values derived from observation for the Milky Way (\citealt{2010ApJ...710L..11R,2011AJ....142..197C}). Middle panel: Evolution of the mass of a stable isotope present in the interstellar medium, assuming different star formation histories. Bottom panel: Evolution of the mass ratio between the radioactive and stable isotopes.  The thick orange lines represent the analytic solutions of the Clayton's model.  All other lines have been computed using a simplified version of \omegap\ that mimics the conditions adopted in Clayton's model (Section~\ref{sec:num_setup}). In all calculations, we assumed a radioactive-to-stable mass ratio of 0.2 for the yields and a half-life of 10\,Myr for the radioactive isotopes.
\label{fig:SFH}}
\end{figure}

Figure~\ref{fig:SFH} shows the time evolution in our Milky Way model of the SFH, the mass of a stable isotope present in the ISM, and the isotopic ratio between a radioactive and a stable isotope. The different lines represent different options for the gas inflow prescription used to generate the SFH (see next paragraph).  All SFHs shown in Figure~\ref{fig:SFH} form the same amount of stars once integrated over the lifetime of the Galaxy. We obtain a final stellar mass of $5.5\times10^{10}$\,M$_\odot$, or $3.5\times10^{10}$\,M$_\odot$ once corrected for the mass returned into the ISM by stellar ejecta. This result is consistent with the $\sim$\,$5\times10^{10}$\,M$_\odot$ derived by \cite{2006MNRAS.372.1149F} using the mass-to-light ratio of the Milky Way, the $(5.17\pm1.11)\times10^{10}$\,M$_\odot$ derived by \cite{2015ApJ...806...96L} for the disk using statistical methods, and the $(3.5\pm1)\times10^{10}$\,M$_\odot$ found in
\cite{2016ARA&A..54..529B} for the thin disk.

The gas inflow rates in Clayton's model are defined as
\begin{equation}
    \dot{M}_\mathrm{inflow}(t) = \frac{k}{t+\Delta}\,M_\mathrm{gas}(t),
\end{equation}
where $k$ and $\Delta$ are free parameters. For the solid, dashed, and dotted black lines in Figure~\ref{fig:SFH}, we used this prescription with $k=0$, 1, and 2, respectively, along with $\Delta=0.5$\,Gyr. We note that using different $\Delta$ values can change the overall shape of the SFH (Appendix~\ref{ap_1}). For the red line, we used the two-infall prescription described in \cite{1997ApJ...477..765C}. This combines two exponential gas inflow episodes defined by
\begin{equation}
    \label{eq_two_infall}
    \dot{M}_\mathrm{inflow}(t) = A_1\,\mathrm{exp}\left(\frac{-t}{\tau_1}\right) + A_2\,\mathrm{exp}\left(\frac{t_\mathrm{max}-t}{\tau_2}\right),
\end{equation}
where $A_1$, $A_2$, $\tau_1$, $\tau_2$, and $t_\mathrm{max}$ are free parameters. Here we set $\tau_1$, $\tau_2$, and $t_\mathrm{max}$ to 0.8, 7.0, and 1.0\,Gyr, respectively, but we left $A_1$ and $A_2$ as free parameters. We also included a constant SFH for completeness to better visualize the impact of using different shapes for the SFH. All models have been adjusted to have the same mass of gas at the end of the simulation in order to isolate the impact of the SFH and to provide a consistent comparison between models.  This has been done by tuning the amount of gas inflow and the star formation efficiency of each model.  The impact of varying the mass of gas and the gas-to-star mass ratio is presented in Section~\ref{sec:sfe}.

The evolution of the isotopic ratio depends on the overall shape (temporal profile) of the SFH of the Galaxy.  The more the SFH peaks at early times (top panel of Figure~\ref{fig:SFH}), the lower will be the $M_\mathrm{radio}/M_\mathrm{stable}$ ratio at the time the Sun forms ($t_\odot\sim8.5$\,Gyr, lower panel of Figure~\ref{fig:SFH}).  This results from three different factors.  First, the mass of the stable isotope is related to the integration of the SFH.  The steeper the SFH is, the more stars will form by time $t_\odot$, and the larger will be $M_\mathrm{stable}$.  Second, the mass of the radioactive isotope at time $t_\odot$ only depends on the star formation rate (SFR) at that time.  Indeed, because the Galactic age is significantly larger than the half-life of the radioactive isotopes, most of the radioactive isotopes ejected at earlier times will have decayed.  Therefore, a steeper SFH implies a lower the SFR at time $t_\odot$ and a smaller $M_\mathrm{radio}$ at that time, which in turn decreases the $M_\mathrm{radio}/M_\mathrm{stable}$ ratio.

The third factor is the fraction of stable isotopes locked inside stars and remnants (see also Section~\ref{sec:sfe} for further explanations).  As shown in the middle panel of Figure~\ref{fig:SFH}, the mass $M_\mathrm{stable}$ present in the interstellar gas at the end of the simulation is higher when the SFH is steeper.  We note that all of our models have produced the same amount of stable isotopes by the end of the simulation.  The variations seen in this middle panel are only caused by variations in the mass of stable isotopes locked away.  A steeper SFH therefore reduces the isotopic ratios because a lower fraction of $M_\mathrm{stable}$ is locked inside stars and remnants. In other words, more stable isotopes are present in form of interstellar gas. 
 
To summarize, when most of the stars form at early times, the three factors described above all contribute to reduce the $M_\mathrm{radio}/M_\mathrm{stable}$ ratio by the time the Sun forms. This is why, in Figure~\ref{fig:SFH}, the steepest SFH has the lowest isotopic ratio, while the flat SFH has the largest one.  As a final note, once a specific shape has been adopted for the SFH, the normalization (total stellar mass formed) does not change the isotopic composition.  Adopting higher or lower SFRs will increase or reduce the total mass of isotopes ejected into the ISM, but will not modify the isotopic ratios, as long as the gas-to-star mass ratio remains the same.  Indeed, as described in Section~\ref{sec:sfe}, assuming different gas-to-star mass ratio does change the evolution of $M_\mathrm{radio}/M_\mathrm{stable}$.

Using the same simplifications as in Clayton's model (Section~\ref{sec:num_setup}), the predictions of our models are exactly the same as the analytical solutions of Clayton's model (bottom panel of Figure~\ref{fig:SFH}). This comparison confirms that the radioactivity implementation in our chemical evolution model works properly. In the next sections, we use the model with the two-infall inflow prescription as the fiducial model.

\subsection{Gas-to-Star Mass Fraction}
\label{sec:sfe}
\begin{figure}
\center
\includegraphics[width=3.35in]{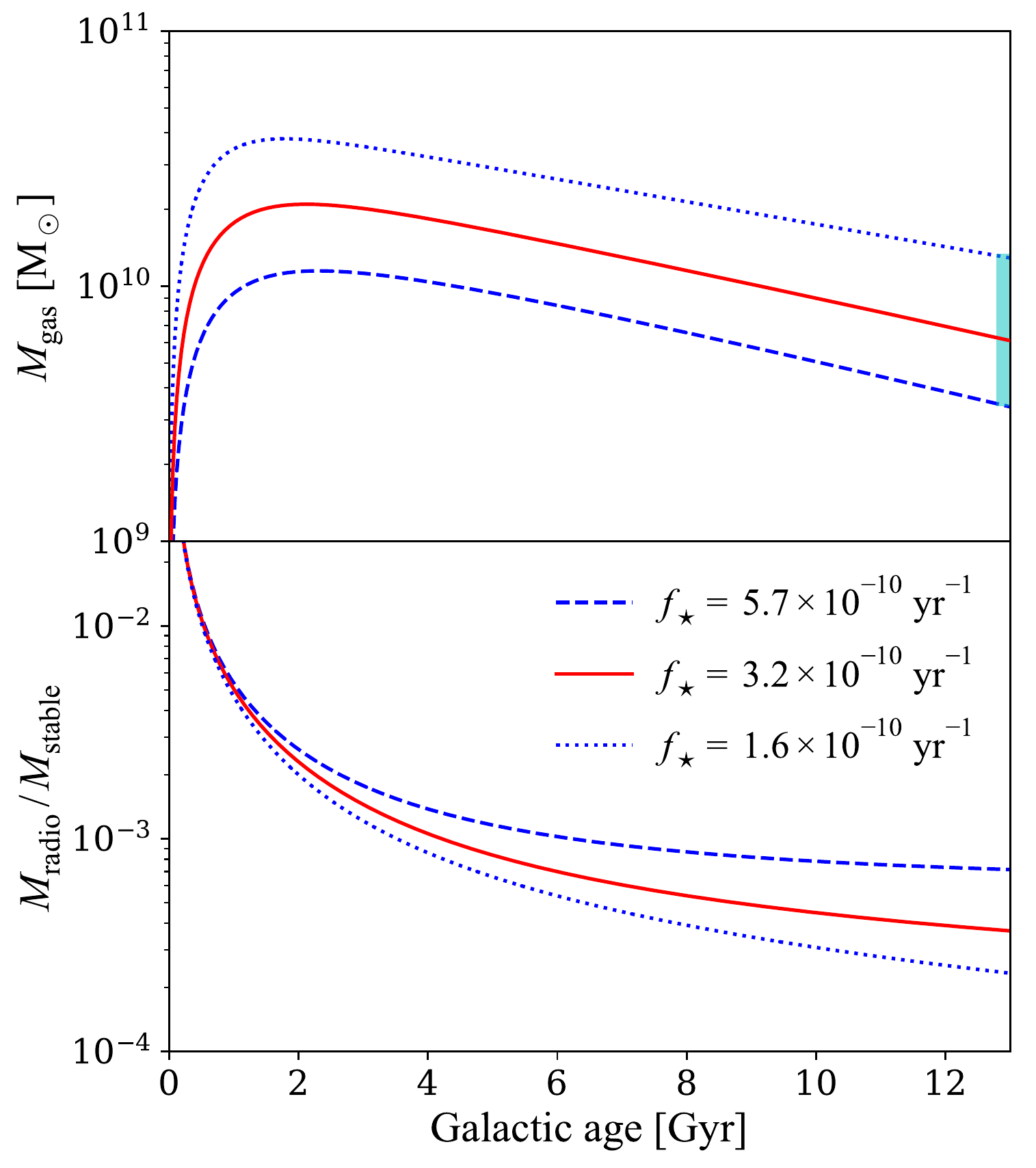}
\caption{Top panel: Evolution of the mass of gas inside the galaxy, using the two-infall prescription of \cite{1997ApJ...477..765C} for the star formation history (red lines in the top panel of Figure~\ref{fig:SFH}). Different lines represent different star formation efficiency (see the $f_\star$ parameter in Equation~\ref{eq:sfe}). The vertical thick cyan line represents the current value derived from observation for the Milky Way (\citealt{2015A&A...580A.126K}). Bottom panel: Evolution of the mass ratio between the radioactive and stable isotopes, for different star formation efficiency.  
\label{fig:sfe}}
\end{figure}

In this section, we explore the impact of varying the gas-to-star mass ratio in the Galaxy, using the two-infall prescription to generate the SFH of our models.  For this experiment, we tuned the magnitude of the inflow rates and the star formation efficiency of each model so that they all generate a similar SFH as in our fiducial case (the red line in the top panel of Figure~\ref{fig:SFH}).  Since all models form the same total stellar mass by the end of the simulation, varying the star formation efficiency only changes the mass of the gas reservoir (ISM) in which stars form and return their ejecta (top panel of Figure~\ref{fig:sfe}).  We set the range of star formation efficiencies in order to reproduce the estimated mass of gas present in the Galactic disk, which ranges from $3.6\times10^9$ to $1.3\times10^{10}$\,M$_\odot$ (\citealt{2015A&A...580A.126K}).  The latter values are also consistent with the observed star formation efficiency of nearby spiral galaxies (\citealt{2008AJ....136.2782L}).

We note that two of the three models presented in this section have gas inflow rates that are too low compared to the value  derived for the Milky Way.  Those inflow rates could be increased by accounting for galactic outflows, without changing the mass of gas and the SFH (Section~\ref{sec:outflows}). However, we do not apply such a correction because our goal is to explain, step by step, the impact of different galaxy evolution processes on the predicted isotopic ratio. We present our final models tuned to respect simultaneously the various observational constraints, including the gas inflow rate and the gas-to-star fraction in Section~\ref{sec:uncertainty}.

As shown in the bottom panel of Figure~\ref{fig:sfe}, adopting a lower star formation efficiency generates a larger gas reservoir and decreases the $M_\mathrm{radio}/M_\mathrm{stable}$ ratio. To understand this, we remind that the same amount of stars is formed in all models. This means that the same mass of isotopes is produced and ejected throughout the simulations.  Because there is no galactic outflow included in this section, all the isotopes produced are therefore either found in the interstellar gas or locked inside stars and remnants.  When increasing the mass of gas (that is, lowering the star formation efficiency), the concentration of stable isotopes is more diluted and therefore a smaller fraction of stable isotopes is locked into stars.  This increases $M_\mathrm{stable}$ and thus reduces the $M_\mathrm{radio}/M_\mathrm{stable}$ ratio. The mass of radioactive isotopes is less affected by the mass of gas. As described in Section~\ref{sec:SFH}, $M_\mathrm{radio}$ at a given time mostly depends the SFR at that time, which is similar from one model to another.

\subsection{Galactic Outflows}
\label{sec:outflows}
\begin{figure}
\center
\includegraphics[width=3.35in]{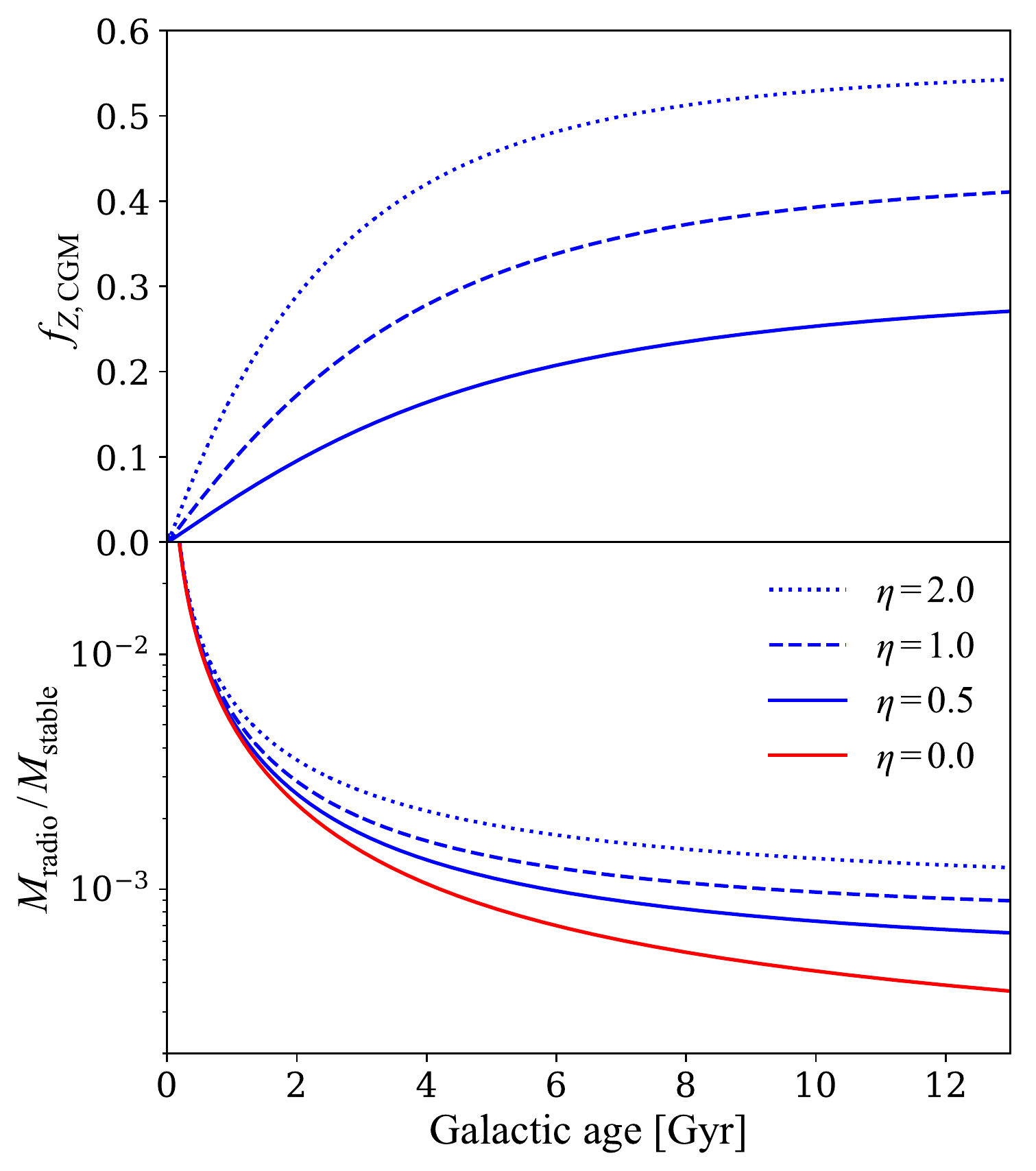}
\caption{Top panel: Evolution of the fraction of  metals, ejected by stars, found outside the galaxy (into the circumgalactic medium, CGM), using the two-infall prescription of \cite{1997ApJ...477..765C} for the star formation history (red lines in the top panel of Figure~\ref{fig:SFH}).  Different lines represent different strength of galactic outflows (see the $\eta$ parameter in Equation~\ref{eq:eta}).  Bottom panel: Evolution of the mass ratio between the radioactive and stable isotopes, for different strength of galactic outflows.  The adopted yields and half-life are the same as in Figure~\ref{fig:SFH}.
\label{fig:eta}}
\end{figure}

In this section, we explore the impact of including and varying the strength of galactic outflows, which remove gas from the galaxy.  All models have the same SFH and the same mass of interstellar gas throughout the simulations (the red lines in the top panels of Figures~\ref{fig:sfe} and \ref{fig:SFH}). To make this calibration, we tuned the intensity of inflows to balance the outflows so that the net amount of mass gained by the galaxy is the same in each model.  The strength of a galactic outflow is defined by the mass-loading parameter $\eta$ (Equation~\ref{eq:eta}).  Figure~\ref{fig:eta} compares our fiducial case ($\eta=0$) with three models that used $\eta=0.5$, 1, and 2, respectively.  These values, of the order of unity, are consistent with the mass-loading factors predicted by cosmological hydrodynamic simulations of Milky Way-like galaxies at low redshifts (e.g., \citealt{2014MNRAS.443.3809B,2015MNRAS.454.2691M,2017MNRAS.470.4698A,2018MNRAS.473.4077P}).

 As shown in the bottom panel of Figure~\ref{fig:eta}, models with stronger galactic outflows (higher $\eta$) show a higher $M_\mathrm{radio}/M_\mathrm{stable}$ ratio.  This is because outflows eject stable isotopes into the CGM, outside the galaxy.
 Although isotopes ejected outside the galaxy can fall back onto the galaxy and be recycled at later times, a significant fraction of stable isotopes is continuously trapped in the CGM (see top panel of Figure~\ref{fig:eta}). As a matter of fact, the COS-Halos Survey (\citealt{2014ApJ...792....8W}) revealed that potentially more than half of all metals produced by stars should be outside galaxies, even for Milky Way-like galaxies (\citealt{2014ApJ...786...54P,2017ARA&A..55..389T}). This fraction is also consistent with the predictions from hydrodynamic galaxy simulations (e.g., \citealt{2016MNRAS.460.2157O,2018arXiv180807872C}). The exact fraction of metals locked outside the Milky Way is difficult to extract given the large uncertainties, but there are clear observational evidences that there is a hot gas reservoir with metals currently surrounding the Milky Way (e.g., \citealt{2016ARA&A..54..529B}).
 
\subsection{Delay-Time Distributions}
\label{sec:DTD}
\begin{figure}
\center
\includegraphics[width=3.35in]{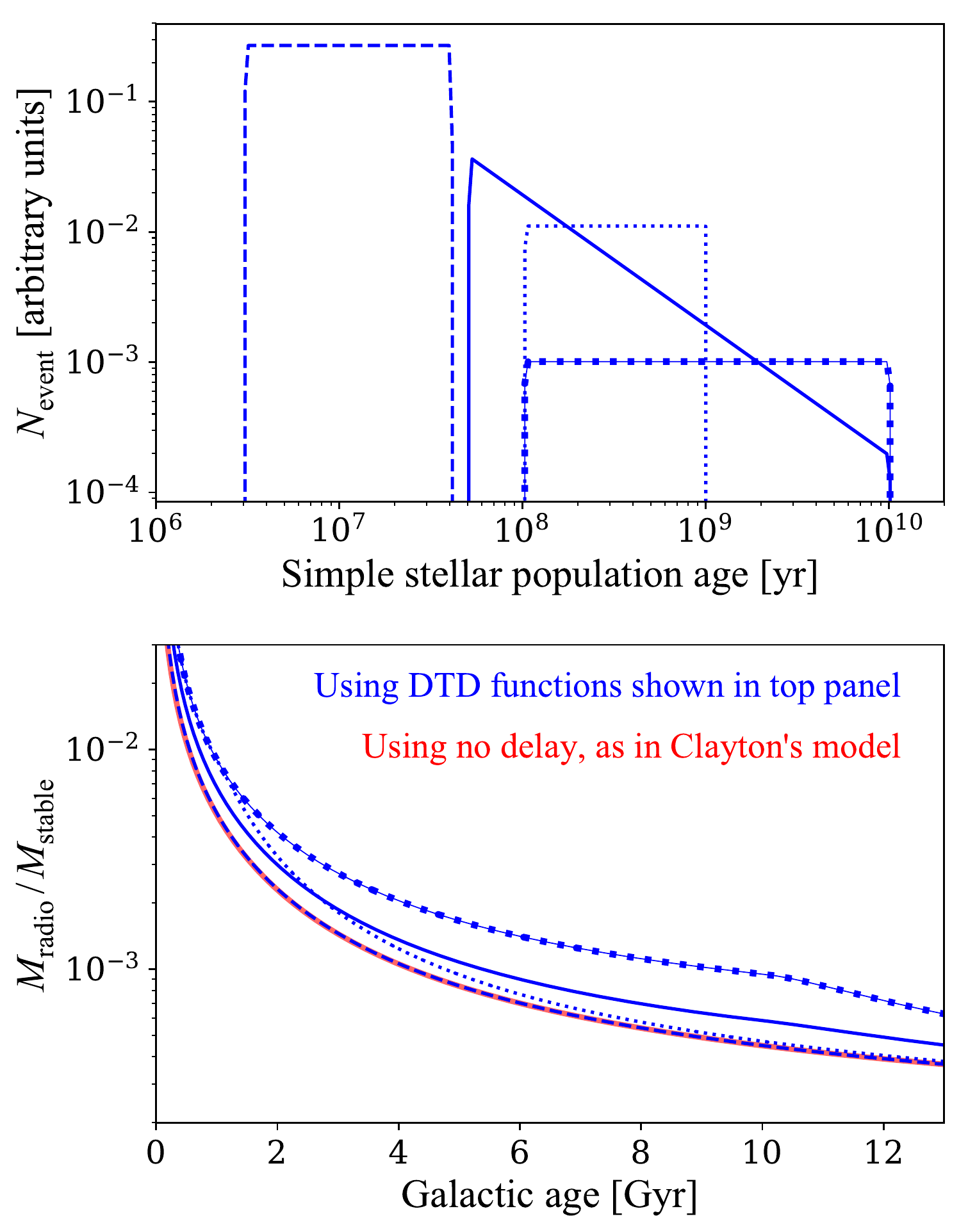}
\caption{Top panel: Examples of delay-time distribution functions that can be associated with the astronomical events producing the stable and radioactive isotopes.  They show how the stellar ejecta are distributed as a function of time in a simple stellar population.  
Bottom panel: Evolution of the mass ratio between the radioactive and stable isotopes, assuming different delay-time distribution functions.  The red line is our fiducial model (red lines in the top panels of Figures~\ref{fig:SFH} and \ref{fig:sfe}), which does not include any delay between the ejection of isotopes and the formation of the progenitor stars, as in Clayton's model.
\label{fig:dtd}}
\end{figure}

In Clayton's model, there is no delay between the stellar ejecta and the formation of their progenitor stars.  In this section, we relax this assumption and study the impact of using different delay-time distribution functions to distribute the stellar ejecta of each stellar population formed throughout the lifetime of our simulated galaxy.  These functions are shown in the top panel of Figure~\ref{fig:dtd}.  While they are only illustrative arbitrary cases, the dashed line can be associated with core-collapse supernovae from massive stars, the solid line can be associated with Type~Ia supernovae or compact binary mergers, and the thin dotted line to stars with initial mass roughly between 2 and 4\,M$_{\odot}$. 

\begin{figure*}
\includegraphics[width=6.9in]{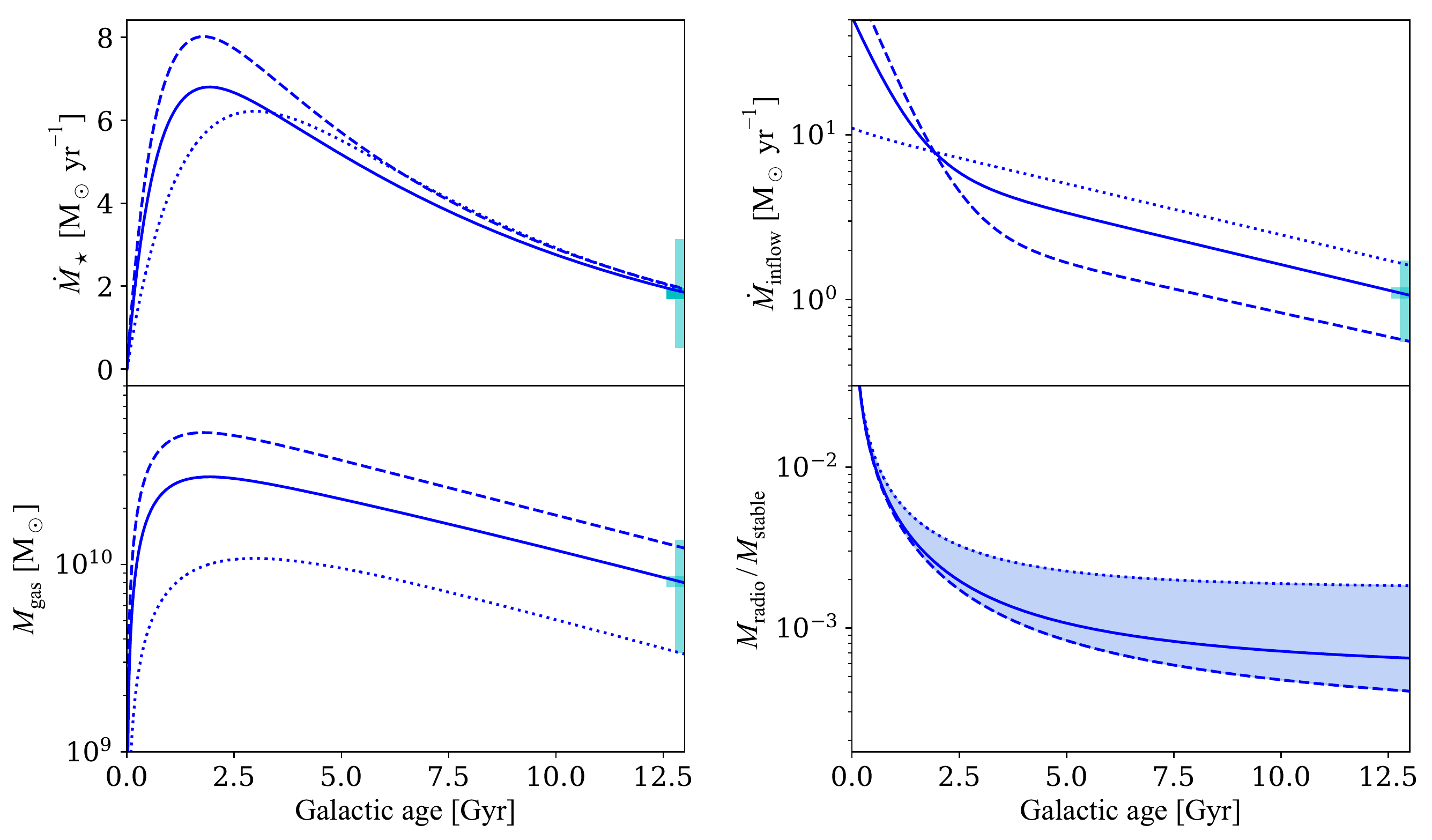}
\caption{Evolution of the star formation rate (top-left panel), gas inflow rate (top-right panel), mass of gas (bottom-left panel), and isotopic mass ratio predicted by our best-fit Milky Way model (blue solid line) and our two extreme models (blue dashed and dotted lines). The cyan bands at 13~Gyr are observational constraints taken from \cite{2015A&A...580A.126K}. The small horizontal thick line within those cyan bands represents the middle point of the interval. 
\label{fig:final}}
\end{figure*}

As shown in the bottom panel of Figure~\ref{fig:dtd}, accounting for delays between the formation of stars and their ejecta can increase the $M_\mathrm{radio}/M_\mathrm{stable}$ ratio. The more the ejecta is concentrated at late times, the larger will be the isotopic ratio.  Indeed, when assuming large delay times of the order of several Gyr, there will be less stable isotopes present in the ISM at a given time, since not all isotopes will have been ejected by that time.  This systematically reduces the accumulated mass $M_\mathrm{stable}$. The shape of the SFH does not play a significant role in the variations seen in the bottom panel of Figure~\ref{fig:dtd}.  When using a constant SFH instead an exponential decreasing SFH, the variations seen in the $M_\mathrm{radio}/M_\mathrm{stable}$ ratio are similar.

Overall, unless the adopted astronomical event has a delay-time distribution function that strongly favours large delay times of the order of several Gyr (e.g., thick dotted line in Figure~\ref{fig:dtd}), the results are not significantly affected.  Indeed, the model that includes delay times similar to the lifetime of massive stars (dashed line) is almost perfectly overlapping the fiducial model (red line).

\subsection{Best-Fit Model and Range of Solutions}
\label{sec:uncertainty}
\begin{figure}
\center
\includegraphics[width=3.35in]{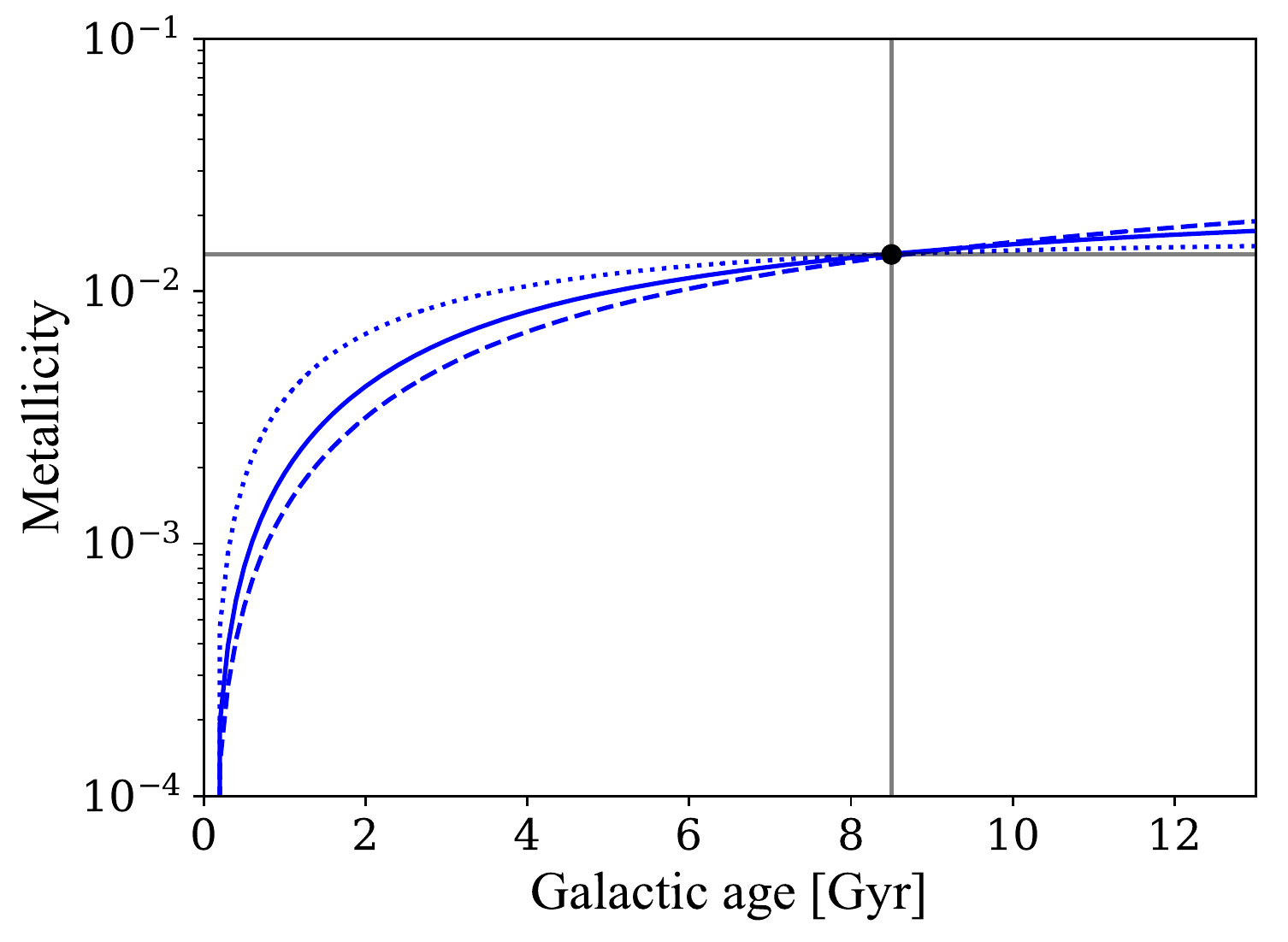}
\caption{Evolution of the gas metallicity (mass fraction) predicted by the three Milky Way models presented in Figure~\ref{fig:final}. The vertical and horizontal gray solid lines represent the time at which the Sun is assumed to form, and the Solar metallicity ($Z=0.014$, \citealt{2009ARA&A..47..481A}), respectively.
\label{fig:Z_vs_t}}
\end{figure}

In the previous sections, we presented how the isotopic ratio can be altered by the shape of the SFH, the gas-to-star mass ratio, the presence of galactic outflows, and the delay-time distribution of the enrichment events.  The goal was to better understand the role played by these basic ingredients, individually.  In Figure~\ref{fig:final} we present our best-fit model\footnote{\url{https://github.com/becot85/JINAPyCEE/blob/master/DOC/OMEGA\%2B_Milky_Way_model.ipynb}} tuned to reproduce simultaneously the following observational constraints for the Milky Way disk: current SFR, gas inflow rate, mass of gas, core-collapse and Type~Ia supernova rates, and total stellar mass formed. Since the observational constraints used to calibrate our Milky Way model have uncertainties, we also present the two extreme models that illustrate the largest variations we can achieve while still remaining within the observational error bars. These extreme models are used to define the confidence level of our isotopic ratio predictions (see also \citealt{dauphas03}).  All three models reach solar metallicity (\citealt{2009ARA&A..47..481A}) by time $t_\odot$ (Figure~\ref{fig:Z_vs_t}).  The level of uncertainties shown in Figure~\ref{fig:final} can be applied to any radioactive isotope with a half-life below $\sim$\,200\,Myr, such as $^{26}$Al and $^{60}$Fe. For longer-lived isotopes such as $^{238}$U and $^{232}$Th, the uncertainty is likely to decrease (see Section~\ref{sec:mod_ss} for discussion).

The parameters and final properties of our models are shown in Table~\ref{tab_final}. We did not include any delay-time distribution, as we want our results to be as general as possible.  Depending on the adopted enrichment source, a shift in the predictions should be included following the results presented in Figure~\ref{fig:dtd} (Section~\ref{sec:DTD}). To generate the SFH, we used the two-infall prescription described in \cite{1997ApJ...477..765C}. We remind that using different prescriptions could shift the results presented in this section (Section~\ref{sec:SFH}). An iPython Jupyter notebook describing how to run \omegap\ using different gas inflow and star formation histories is available on the JINAPyCEE GitHub repository\footnote{\url{https://github.com/becot85/JINAPyCEE/blob/master/DOC/OMEGA\%2B_defining_gas_inflow.ipynb}} for further explorations.

\subsubsection{Minimizing the Isotopic Ratio}

The lowest $M_\mathrm{radio}/M_\mathrm{stable}$ ratio in Figure~\ref{fig:final} was obtained by steepening the slope of the SFH, relative to that of the best-fit model. This was done by increasing the magnitude of the first gas infall episode and by decreasing the magnitude of the second one. As described in Section~\ref{sec:SFH}, the more the SFH peaks at early time, the more a stable isotope is produced by time $t_\odot$. We also increased the total stellar mass formed to maximize the production of stable isotopes. In practical terms, we reduced the second infall until we reached the lower limit for the observed galactic inflow rate (top-right panel of Figure~\ref{fig:final}), and we increased the first infall until we reached the upper limit for the observed stellar mass. 

To further minimize the $M_\mathrm{radio}/M_\mathrm{stable}$ ratio, we decreased the star formation efficiency to increase the gas-to-star ratio. As described in Section~\ref{sec:sfe}, for the same stellar mass formed, more gas inside the galaxy minimizes the amount of stable isotopes locked into stars and remnant, which in turn maximizes $M_\mathrm{stable}$. In practical terms, since decreasing the star formation efficiency also decreases the total stellar mass formed, we further increased the magnitude of the first gas infall episode to maintain the same total stellar mass. 

Shutting down galactic outflow should in theory help minimizing the isotopic ratio (Section~\ref{sec:outflows}). But as shown in Table~\ref{tab_final}, all models have outflows with a mass-loading factor of $\sim$\,0.5. This value ensured to reach solar metallicity by $t_\odot$ (Figure~\ref{fig:Z_vs_t}). If outflows were removed from the minimizing model, the metallicity of the gas would be too high and the mass of gas would increase beyond the upper limit set by observations. One way to reduce the metallicity would be to decrease the star formation efficiency. But doing so would further increase the mass of gas. To decrease the mass of gas without outflow, the inflow rate could be decreased. But doing so would decrease the current inflow rate below the lower limit set by observations. We note that with $\eta\sim0.5$, about 25\,\% of all metals produced in our simulations reside outside the galaxy (Figure~\ref{fig:eta}, see also \citealt{2012MNRAS.425.1270S}).
\\
\subsubsection{Maximizing the Isotopic Ratio}

The opposite operations have been done to obtain the highest possible $M_\mathrm{radio}/M_\mathrm{stable}$ ratio. In particular, the first gas infall episode has been practically removed to minimize the stellar mass formed, and the star formation efficiency has been increased to minimize the mass of gas. As mentioned above, we did not have much room to vary the strength of galactic outflows. In theory, having more outflows should increase the isotopic ratio. But with more outflows, the total stellar mass formed would decrease below the lower limit set by observations. Increasing the star formation efficiency to increase the stellar mass would lower the current mass of gas below the lower limit. Increasing the inflow rate to increase the mass of gas would increase the current inflow rate beyond the observed upper limit.


\begin{deluxetable}{lllll}
\tablewidth{0pc}
\tablecaption{Parameters (top segment) and final properties (bottom segment) of the Milky Way models shown in Figure~\ref{fig:final}. The Low and High models provide the lowest and highest isotopic ratios ($M_\mathrm{radio}/M_\mathrm{stable}$) and are shown as dashed and dotted lines in Figure~\ref{fig:final}, respectively. The input parameters are the normalization of the first and second infall episodes ($A_1$ and $A_2$, Equation~\ref{eq_two_infall}), star formation efficiency ($f_\star$, Equation~\ref{eq:sfe}) and strength of galactic outflows ($\eta$, Equation~\ref{eq:eta}). The final properties are the current gas inflow rate ($\dot{M}_{\mathrm{inflow},0}$), mass of gas ($M_{\mathrm{gas},0}$), star formation rate ($\dot{M}_{\star,0}$), mass of stars ($M_{\star,0}$), and core-collapse ($R_{\mathrm{CC},0}$) and Type~Ia ($R_{\mathrm{Ia},0}$) supernova rates. The observational constraints are taken from the compilation found in Kubryk et al.\,(2015).
\label{tab_final}}
\tablehead{
 \multirow{2}{*}{Quantity} & \multicolumn{3}{c}{Milky Way Models} & \multirow{2}{*}{Observations} \\ 
 \colhead{ } & \colhead{Low} & \colhead{Best} & \colhead{High}
 }
\startdata
$A_1$ [M$_\odot$\,yr$^{-1}$] & 91 & 46 & 0.7 & -- \\
$A_2$ [M$_\odot$\,yr$^{-1}$] & 2.9 & 5.9 & 9.0 & --  \\
$f_\star$ [$10^{-10}$\,yr$^{-1}$] & 1.6 & 2.3 & 5.8 & --  \\
$\eta$ & 0.50 & 0.52 & 0.45 & --  \\
\hline
$\dot{M}_{\mathrm{inflow},0}$ [M$_\odot$\,yr$^{-1}$] & 0.57 & 1.1 & 1.6 & $0.6-1.6$ \\
$M_{\mathrm{gas},0}$ [$10^{10}$\,M$_\odot$] & 1.3 & 0.80 & 0.33 & $0.36-1.3$ \\
$\dot{M}_{\star,0}$ [M$_\odot$\,yr$^{-1}$] & 2.0 & 1.9 & 1.9 & $0.65-3$ \\
$M_{\star,0}$ [$10^{10}$\,M$_\odot$] & 4.1 & 3.6 & 3.4 & $3-4$ \\
$R_{\mathrm{CC},0}$ [century$^{^-1}$] & 1.9 & 1.8 & 1.9 & $1-3$ \\
$R_{\mathrm{Ia},0}$ [century$^{^-1}$] & 0.32 & 0.29 & 0.29 & $0.2-0.6$ \\
\enddata
\end{deluxetable}

\subsubsection{Modified Steady-State Equation}
\label{sec:mod_ss}
The results shown in the bottom panel of Figure~\ref{fig:final} at $t_\odot$ can be recovered also by the steady-state formula. Using Equation~(\ref{eq:steady}) with $T_\mathrm{Gal}=t_{_\odot}=8.5$\,Gyr, a half-life of 10\,Myr ($\tau=14.4$\,Myr), and a stellar production ratio of 0.2 as used in our simulations, our best-fit model is recovered by multiplying the steady-state result by 2.3. The lower and upper limits are recovered by multiplying the result by 1.6 and 5.7, respectively. We repeated the experiment with nine different mean lives from 1 to 200\,Myr in order to test the robustness of this comparison. For mean lives below $\sim20$\,Myr, all of our multiplication factors are the same. For longer mean lives, the factors slightly decrease. At $\sim200$\,Myr, the upper, best-fit, and lower values stated above decreased by 12\,\%, 7\,\%, and 4\,\%, respectively. When targeting long-lived isotopes such as $^{238}$U and $^{232}$Th, we thus recommend to use our codes instead of using the multiplication factors mentioned above. The width of the uncertainty band is not affected by the choice of the production ratio, but the absolute value of the isotope ratio is directly proportional to that choice.

Our best-fit model is consistent with the multiplication factor of $2.7\pm0.4$ calculated by \citet{dauphas03} using their analytical model. However, the range we obtain is wider than that of \citet{dauphas03}. This is likely because the error bars associated with the observations used in our work are larger than those used in \citet{dauphas03}. We remind that the multiplication factors derived in this section and in \cite{dauphas03} do not account for the effect of the delay-times distribution of the consider source (Section~\ref{sec:DTD}). If the adopted enrichment source has long delay times such as Type~Ia supernovae or low-mass asymptotic giant branch stars, the multiplications factors should be increased (see Figure~\ref{fig:dtd}).


\section{Discussion}
\label{sec:disc}
In this section, we discuss the uncertainties in our predictions and highlight the role of our numerical framework in studying the conditions that led to the formation of the Solar System.

\subsection{Level of Uncertainties}
\label{sec:disc_uncertainties}
As discussed above, when the target isotopic ratio involves a short-lived ($\sim$\,Myr) radioactive and a stable isotope, the predicted isotopic composition of the ISM at the time the Sun formed is uncertain by a factor of 3.6 (blue shaded area in Figure~\ref{fig:final}). This represents the maximum level of uncertainty, given the number of uncertainty sources included in our models (Section~\ref{sec:results}).  A better way to quantify the output uncertainties of our GCE models would be to calculate a large number of models where the input parameters would be randomly selected before each run, in a Monte Carlo fashion (see e.g., \citealt{2016ApJ...824...82C}). This would provide the probability distribution function of the predicted ratios, instead of a flat uncertainty band as shown here. This will be explored in further studies.

\begin{figure}
\includegraphics[width=3.35in]{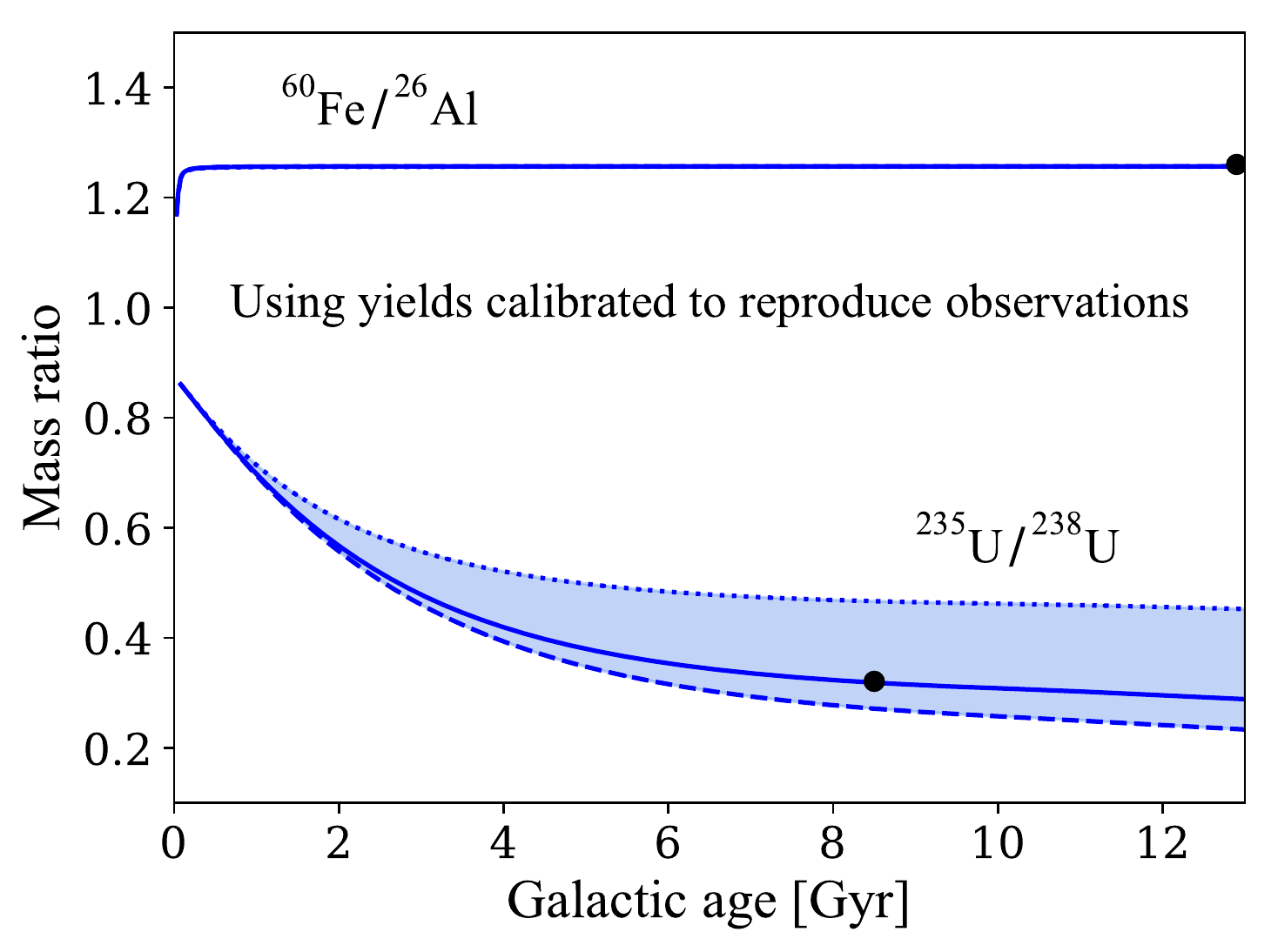}
\caption{Evolution of $^{60}$Fe\,/\,$^{26}$Al and $^{235}$U\,/\,$^{238}$U predicted by our homogenized Milky Way model. We used arbitrary yields calibrated to reproduce the $^{60}$Fe\,/\,$^{26}$Al ratio currently observed in the interstellar medium \citep{wang07} and the $^{235}$U\,/\,$^{238}$U ratio inferred for early Solar System using meteorite data analysis \citep{lodders10}. The lines are the same as in Figure~\ref{fig:final}.
\label{fig:fe_al_u}}
\end{figure}

We remind that the mass of radioactive isotopes in our models only depends on the value of the star formation rate, while the mass of stable isotopes probes the total integrated amount of stable isotopes produced throughout the history of the Milky Way. The level of uncertainty is significantly reduced when the stable isotope in the $M_\mathrm{radio}/M_\mathrm{stable}$ ratio is replaced by another radioactive isotope. Overall, the shorter-lived the radioactive isotopes are, the less they are affected by the galaxy evolution uncertainties explored in this work. As an example, Figure~\ref{fig:fe_al_u} shows the evolution of the ratios of two pairs of radioactive isotopes in our Milky Way model. $^{235}$U and $^{238}$U are long-lived isotopes with a half-life of 0.7 and 4.5\,Gyr, respectively. Although $^{238}$U has more memory of the past production of uranium than $^{235}$U, none of them carries the complete production history since the formation of the Galaxy. As a result, by the time the Sun formed, the predicted $^{235}$U\,/\,$^{238}$U ratio is only uncertain by  $\sim$\,60\,\%, as opposed to a factor of 3.6. When following the evolution of two very short-lived radioactive isotopes, such as $^{60}$Fe\,/\,$^{26}$Al, with half-lives of 2.6 and 0.72\,Myr, respectively, galaxy evolution uncertainties do not have any impact as their abundances do not carry any trace of past nucleosynthesis production. We note that to generate the predictions shown in Figure~\ref{fig:fe_al_u}, we assumed that the production ratios in the yields were constant throughout our GCE calculations, and used arbitrary yields tuned to reproduced the observed $^{60}$Fe\,/\,$^{26}$Al and $^{235}$U\,/\,$^{238}$U ratios. In future studies, however, our codes will enable to use theoretical nucleosynthesis yields to properly follow the production of radioactive isotopes (Section~\ref{sec:disc_future}).

Galaxy evolution uncertainties therefore do not always affect ratios involving radioactive isotopes. In the case of $^{60}$Fe\,/\,$^{26}$Al, within the continuous and homogenized enrichment approximation, the observations directly probe nuclear astrophysics and the nucleosynthesis of $^{60}$Fe and $^{26}$Al in stellar environments, with no effect from galaxy evolution uncertainties. On the other hand, the ratios $M_\mathrm{radio}/M_\mathrm{stable}$ involving a stable isotope are significantly affected by those uncertainties (Figure~\ref{fig:final}). Using such ratios to constrain and probe nuclear astrophysics becomes more challenging, as galaxy evolution and nuclear astrophysics uncertainties could alter the predicted ratios by similar amounts. Our uncertainties represent only those deriving from GCE. In this work we did not include nuclear physics uncertainties such as the error bars on the half-lives, nor stellar yields uncertainties.

Stellar uncertainties can affect in particular the predicted $M_\mathrm{radio}/M_\mathrm{stable}$ ratios, if isotopes are made by different nucleosynthesis processes and/or at different conditions. For instance, in the $^{60}$Fe/$^{56}$Fe ratio, $^{60}$Fe is mostly a neutron capture product, while the bulk of $^{56}$Fe is made as $^{56}$Ni in extreme supernovae conditions. In the $^{26}$Al/$^{27}$Al ratio, $^{27}$Al is efficiently made by neutron capture on $^{26}$Mg, while $^{26}$Al is partially destroyed by (n,p) and (n,$\alpha$) neutron capture reactions \citep[e.g.,][]{timmes95,limongi:06, sukhbold:16}. Therefore, GCE uncertainties are probably a lower limit on the total uncertainties, although the effect of some of them may cancel each other. Statistical studies are required to qualitatively evaluate these combined effects. 

\subsection{The Role of Our Numerical Framework}
\label{sec:disc_future}

Our GCE codes allow to follow in detail the evolution of radioactive-to-stable isotope ratios in the Galaxy. Compared to using a simple steady-state formula or an analytical model, our framework is more flexible and can easily incorporate new developments from the galaxy evolution community. In addition, mass- and metallicity-dependent stellar yields can be used. To summarize, our framework offers a unique opportunity to reinforce the connections between cosmochemistry, nuclear astrophysics, nucleosynthesis, and galaxy evolution. Another important aspect of our codes is that multiple nucleosynthesis sources contributing to the same isotope can be followed accurately. For example, radioactive isotopes heavier than iron and their reference stable isotopes such as the \iso{107}Pd$-$\iso{108}Pd and \iso{182}Hf$-$\iso{180}Hf pairs are produced both by the rapid and the slow neutron capture processes. While the former behaves in a primary fashion, the latter has a different dependency on metallicity depending if the isotope is located near to the first or the second $s$-process peak \citep[see, e.g.][]{travaglio99,travaglio04}. While we have provided a way to still use the steady-state equation, many cases such as those mentioned above can only be followed accurately with numerical GCE models \citep[see e.g., ][]{travaglio14}. 

The main limitation of the GCE calculations performed in this work is the assumption that the ISM is uniformly mixed. Our predictions should thus be seen as a representation of the average chemical evolution of our Galaxy. Given this limitation, the current version of our codes cannot predict the uncertainties deriving from the effect of chemical inhomogeneities in the ISM at the time of formation of the molecular cloud in which the Sun was born. Neither can it account for the chemical signatures of potential last-injection events within such molecular cloud (e.g., a supernova, a stellar wind) that found their way into the Solar System prior to its formation. Those aspects, however, must be accounted for in order to best interpret the presence of radioactive isotopes in the early Solar System, as inferred from meteorite data analysis. Within this context, our chemical evolution framework is designed to provide the averaged initial chemical composition of the ISM at the time of the formation of the Sun, on top of which follow-up studies \citep[such as those of][]{gaidos09,gounelle12,vasileiadis13,young14,2015A&A...577A.139C,wehmeyer15,hotokezaka15,fujimoto18} could include inhomogeneities and last-injection events to explain some of the signatures seen in meteorites.


As described in detail in \citet{LUGARO2018}, the effect of ISM inhomogeneities is an additional error bar to be added to the radioactive-to-stable isotope ratio at the time of the formation of the Sun. This error bar is a strong function of the ratio $\tau/\delta$, where $\tau$ is the mean life of the radioactive isotope and $\delta$ the recurrence time between the stellar additions of matter from a given production site into a specific portion of the ISM. If $\tau/\delta < 0.1$, the distribution of the radioactive isotope is completely inhomogenous in the ISM (i.e., the radioactive to stable abundance ratio oscillates between 0 and the production ratio), while for $\tau/\delta > 10$, the distribution is homogeneous within 10\,\%. Because we do not have a clear understanding of the value of $\delta$ for different nucleosynthetic events, and since different types of events can contribute to the same isotope, follow-up studies of transport of nucleosynthetic ejecta in the ISM, such as the work of \cite{fujimoto18}, are needed to address these uncertainties. 

Still, the present framework can be employed to investigate with relative confidence some of the longest living radioactive isotopes that were present in the early Solar System. For example, it could be used to investigate the radioactive isotopes produced by the $p$ process in supernovae: \iso{146}Sm ($\tau$ of the order of 100-150\,Myr) and \iso{92}Nb ($\tau$ of 50\,Myr). The recurrence time $\delta$ of their production events is likely to be much lower than their mean lives \citep[see e.g.,][]{travaglio14}. Also the radioactive isotope produced by the $r$ process \iso{244}Pu has a relatively long mean life ($\tau$=115 Myr). However, if \iso{244}Pu originates from neutron star mergers, then its recurrence time $\delta$ may be similar or even longer than its mean life \citep[see discussion in][]{LUGARO2018}. The longer living ($\sim$\,Gyr) isotopes of U and Th may be potential test cases. The mean lives of the $s$-process radiaoctive isotopes \iso{107}Pd, \iso{182}Hf, and \iso{205}Pb are of the order of 10-20\,Myr, which may be comparable to the recurrence time of their $s$-process production events, asymptotic giant branch stars with initial masses in between $\sim$ 1.5 and 4\,M$_\odot$.

We note that although our code includes a circumgalactic gas component, it does not include the contribution of a stellar halo component, as in the GCE code of \cite{travaglio04,travaglio14}. This, however, should not impact our predictions at the time of the formation of the Solar System, as Galactic halo stars only represent $\sim$\,1\,\% of the total stellar mass found in the disk (see, e.g., \citealt{2016ARA&A..54..529B}).

\subsection{Short-Lived Radioactive Nuclei in the Early Solar System (ESS)}
\label{sec:disc_ESS}

In Tables~\ref{tab:slr2} and \ref{tab:slr1} we apply both Equation~\ref{eq:steady} with our recommended multiplication factors and the full GCE code to the short-lived radioactive nuclei whose ESS abundances are well determined \citep[according to Table 2 of][]{LUGARO2018}, plus $^{60}$Fe, which is particularly interesting given its $\gamma$-ray detection. We calculate their ratio, with respect to the given reference isotope, in the ISM at the Galactic time of the formation of the Sun, and by applying a free decay between this value and the ESS value we obtain the isolation times reported in the tables. The error bars on the ESS abundances are not shown here as they are small enough to not have any significant effect on the isolation times. For this exercise, we assume constant stellar production ratios, as indicated in the tables and chosen as in \citet{LUGARO2018}, see references and discussion there. When using Equation~\ref{eq:steady}, the production ratios are averaged according to the weights of the different nucleosynthetic sources given in the tables, while in the GCE code the different stellar sources are treated separately and each are given an individual production ratio. The weights of the different sources are estimated based on the contribution of the different processes to the Solar System abundance of the stable isotope of reference. 

\begin{deluxetable*}{lc|cccc}
\tablewidth{0pc}
\tablecaption{Isotopic ratios ($M_\mathrm{radio}$/$M_\mathrm{ref}$) in the ISM when the Sun formed, and isolation times of the pre-solar molecular cloud from the ISM, as predicted by our GCE code and by our analytical approximation (steady-state Equation~\ref{eq:steady} times $2.3^{+3.4}_{-0.7}$), for four radioactive isotopes produced by the $s$ and $r$ processes. Constant production ratios are used for the  yields. When using the GCE code, we assume that the $s$ process takes place in asymptotic giant branch stars with initial mass between 1.5 and 4\,M$_\odot$, which generates a delay-time distribution function in the range from $\sim$\,200\,Myr to $\sim$\,3\,Gyr. For the $r$ process, we assume that it takes place either in rare classes of core-collapse supernovae, or in compact binary mergers with a delay-time distribution function in the form of $t^{-1}$ from 30\,Myr to 10\,Gyr. All isolation times were calculated by finding the time when our ISM isotopic ratio predictions cross the free-decay equation going through the ESS value, using the mean lives of the corresponding radioactive isotopes. 
\label{tab:slr2}}
\tablehead{
\multicolumn{2}{l}{} & \colhead{$^{107}$Pd} & \colhead{$^{129}$I} & \colhead{$^{182}$Hf} & \colhead{$^{247}$Cm}
}
\startdata
\multicolumn{2}{l|}{$\tau$ (Myr)} & 9.4 & 22.6 & 12.8 & 22.5\\
\multicolumn{2}{l|}{Reference isotope} & $^{108}$Pd & $^{127}$I & $^{180}$Hf & $^{235}$U ($\tau \simeq$ 1 Gyr) \\
\multicolumn{2}{l|}{$M_\mathrm{radio}$/$M_\mathrm{ref}$ (ESS)} & 6.6 $\times$ 10$^{-5}$ & 1.28 $\times$ 10$^{-4}$ & 1.02 $\times$ 10$^{-4}$ & 5.6 $\times$ 10$^{-5}$ \\[0.08cm]
\hline
\multicolumn{3}{c}{ } & \multicolumn{2}{c}{Production ratio} & \\[0.05cm]
\hline
\multirow{2}{*}{GCE code$^a$} & $s$ process & 0.14  (65\%) & 0 (5\%) & 0.15 (75\%)& --- \\
& $r$ process & 2.09  (35\%) & 1.35 (95\%) & 0.91 (25\%) & 0.30\\
Equation$^b$ &  & 0.83 & 1.28 & 0.34 & 0.30 \\[0.08cm]
\hline
 \multicolumn{3}{c}{ } & \multicolumn{2}{c}{$M_\mathrm{radio}$/$M_\mathrm{ref}$} & \\[0.08cm]
 \hline
\multirow{3}{*}{GCE code$^c$}
 & Max & [4.89 - 5.45] $\times 10^{-3}$ & [1.93 - 2.15] $\times 10^{-2}$ & [2.83 - 3.07] $\times 10^{-3}$ & [1.18 - 1.17] $\times 10^{-2}$\\
 & Best & [2.02 - 2.37] $\times 10^{-3}$ & [7.74 - 9.46] $\times 10^{-3}$ & [1.18 - 1.32] $\times 10^{-3}$ & [8.13 - 8.52] $\times 10^{-3}$\\
 & Min & [1.43 - 1.73] $\times 10^{-3}$ & [5.36 - 6.93] $\times 10^{-3}$  & [8.37 - 9.63] $\times 10^{-4}$ & [7.32 - 7.73] $\times 10^{-3}$\\[0.15cm]
 \multirow{3}{*}{Equation$^d$}
 & Max & 5.19 $\times 10^{-3}$ & 1.95 $\times 10^{-2}$ & 2.93 $\times 10^{-3}$ & 3.79 $\times 10^{-2}$\\
 & Best & 2.10 $\times 10^{-3}$ & 7.88 $\times 10^{-3}$ & 1.18 $\times 10^{-3}$ & 1.53 $\times 10^{-2}$\\
 & Min & 1.46 $\times 10^{-3}$ & 5.48 $\times 10^{-3}$  & 8.21 $\times 10^{-4}$ & 1.06 $\times 10^{-2}$ \\[0.08cm]
 \hline
 \multicolumn{3}{c}{ } & \multicolumn{2}{c}{Isolation time (Myr)} & \\[0.08cm]
 \hline
\multirow{3}{*}{GCE code$^c$} 
 & Max & [40 - 41]  & [114 - 116] & [43 - 44] & [122 - 123]\\
 & Best & [32 - 34]  & [93 - 98]  & [31 - 33] & [115 - 115]\\
 & Min & [29 - 31]  & [85 - 91]  & [27 - 29] & [112 - 113]\\[0.15cm]
 \multirow{3}{*}{Equation$^d$} 
 & Max & 41 & 114 & 43 & 150\\
 & Best & 32 & 94  & 31 & 129\\
 & Min & 29 & 85  & 27 & 121\\[0.08cm]
\enddata
$^a$ The percentages in parenthesis represent the $s$- and $r$-process contributions to the solar composition of the considered stable reference isotope \citep{arlandini99,bisterzo10}.\\
$^b$ When the $s$- and $r$-process both contribute to the considered isotopes, the equation uses an average production ratio weighted by the percentages shown in parenthesis.\\
$^c$ The values in square brackets show the predictions when assuming that the $r$-process isotopes are produced in rare classes of core-collapse supernovae (values on the left) or in compact binary mergers (values on the right).\\
 $^d$ For $^{247}$Cm/$^{235}$U, we replaced the time variable in Equation~\ref{eq:steady} with the meanlife of $^{235}$U (see Section~\ref{sec:disc_ESS} for discussion).
\end{deluxetable*}

\begin{deluxetable*}{lc|ccccc}
\tablewidth{0pc}
\tablecaption{Same as Table~\ref{tab:slr2}, but for five radioactive isotopes produced in Type~Ia supernovae (SNe~Ia) and core-collapse supernovae (CC~SNe). When using SNe~Ia in the GCE code, we assume either the double-degenerate scenario with a 10-Gyr delay-time distribution in the form of $t^{-1}$, or the single-degenerate scenario with the delay-time distribution predicted by the population synthesis model of \cite{2009ApJ...699.2026R}. The symbol `` --- '' indicates that it is not possible to obtain an isolation time since the ESS ratio is higher than the predicted ISM ratio. \label{tab:slr1}}
\tablehead{
\colhead{ } & \colhead{ } & \colhead{$^{26}$Al} & \colhead{$^{53}$Mn} & \colhead{$^{60}$Fe} & \colhead{$^{92}$Nb} & \colhead{$^{146}$Sm} 
}
\startdata
\multicolumn{2}{l|}{$\tau$ (Myr)} & 1.04 & 5.40 & 3.78 & 50.1 & (98, 149)\\
\multicolumn{2}{l|}{Reference isotope} & \multicolumn{1}{c}{$^{27}$Al} & $^{55}$Mn & $^{56}$Fe & $^{92}$Mo & $^{144}$Sm \\
\multicolumn{2}{l|}{$M_\mathrm{radio}$/$M_\mathrm{ref}$ (ESS)} & 5.23 $\times$ 10$^{-5}$ & 7 $\times$ 10$^{-6}$ & 1.01 $\times$ 10$^{-8}$ & 3.2 $\times$ 10$^{-5}$ & 8.28 $\times$ 10$^{-3}$\\[0.08cm]
\hline
\multicolumn{2}{c}{ } & \multicolumn{5}{c}{Production ratio} \\[0.05cm]
\hline
\multirow{2}{*}{GCE code$^a$} & SNe Ia & --- & 0.108  (60\%) & 0 (70\%) & 1.5 $\times$10$^{-3}$ & 0.35\\
& CC SNe & 4.85 $\times 10^{-3}$ & 0.174 (40\%)& 5.89 $\times 10^{-4} $ (30\%) & --- & ---\\
Equation$^b$ & & 4.85 $\times 10^{-3}$ & 0.134 & 1.76 $\times 10^{-4}$ & 1.5 $\times$10$^{-3}$ & 0.35 \\[0.08cm]
\hline
\multicolumn{2}{c}{ } & \multicolumn{5}{c}{$M_\mathrm{radio}$/$M_\mathrm{ref}$} \\[0.05cm]
\hline
\multirow{3}{*}{GCE code$^{c,d}$}  
 & Max & 3.36 $\times 10^{-6}$ & [5.14 - 5.49] $\times 10^{-4}$ & [4.41 - 4.70] $\times 10^{-7}$ & 6.13 $\times 10^{-5}$ & (2.70, 3.95) $\times$10$^{-2}$\\
 & Best & 1.33 $\times 10^{-6}$ & [2.15 - 2.44] $\times 10^{-4}$ & [1.83 - 2.06] $\times 10^{-7}$  & 3.05 $\times 10^{-5}$ & (1.37, 2.05) $\times$10$^{-2}$\\
 & Min & 9.20 $\times 10^{-7}$ & [1.54 - 1.81] $\times 10^{-4}$ & [1.29 - 1.49] $\times 10^{-7}$ & 2.40 $\times 10^{-5}$ & (1.09, 1.63) $\times$10$^{-2}$\\[0.15cm]
\multirow{3}{*}{Equation$^{d}$}
 & Max & 3.37 $\times 10^{-6}$ & 4.86 $\times 10^{-4}$ & 4.47 $\times 10^{-7}$ & 5.04 $\times 10^{-5}$ & (2.30, 3.49) $\times$10$^{-2}$\\
 & Best & 1.36 $\times 10^{-6}$ & 1.96 $\times 10^{-4}$ & 1.80 $\times 10^{-7}$ & 2.03 $\times 10^{-5}$ & (9.29, 14.1) $\times$10$^{-3}$\\
  & Min & 9.45 $\times 10^{-7}$ & 1.36 $\times 10^{-4}$ & 1.25 $\times 10^{-7}$ & 1.41 $\times 10^{-5}$ & (6.46, 9.79) $\times$10$^{-3}$\\
 \hline
 \multicolumn{2}{c}{ } & \multicolumn{5}{c}{Isolation time (Myr)} \\[0.08cm]
 \hline
 \multirow{3}{*}{GCE code$^{c,d}$}
 & Max & --- & [23 - 24] & [14 - 15] & 33 & (117, 234)\\
 & Best & --- & [19 - 19] & [11 - 11] & ---   & (50, 137)\\
 & Min & --- & [17 - 18] & [10 - 10] & ---   & (27, 103)\\[0.15cm]
\multirow{3}{*}{Equation$^{d}$}
 & Max & --- & 23 & 14 & 23  & (102, 218)\\
 & Best & --- & 18 & 11 & --- & (11, 80)\\
 & Min & --- & 16 & 10 & --- & (---, 25)\\
 \enddata
$^a$ The percentages in parenthesis represent the SNe~Ia and CC~SNe contributions to the Solar composition of the considered stable reference isotope \citep{seitenzahl13,matteucci14}.\\
$^b$ When SNe~Ia and CC~SNe both contribute to the considered isotopes, the equation uses an average production ratio weighted by the percentages shown in parenthesis.\\
$^c$ The values in square brackets show the predictions when assuming the double-degenerate scenario (values on the left) and the single-degenerate scenario (values on the right) for SNe~Ia.\\
$^d$ For $^{146}$Sm, the values in curved parenthesis show the predictions using the two different mean lives reported in the second row of the table \citep{kinoshita12,marks14}. In both cases, we assumed the single-degenerate scenario for SNe~Ia, as in \cite{travaglio14}.
\end{deluxetable*}

For the $r$ process, when using the GCE code we tested both an origin from massive stars and from neutron star mergers. For $^{107}$Pd, $^{129}$I, and $^{182}$Hf, the results obtained within the massive stars framework are equivalent to using Equation~\ref{eq:steady}. With neutron star mergers, the isotopic ratios are higher because of the longer delay times, which leads to slightly longer isolation times. For $^{247}$Cm, on the other hand, the results always differ between the code and the equation. This is because the reference isotope of $^{247}$Cm, $^{235}$U, is also unstable. In principle, Equation~\ref{eq:steady} can be applied to calculate the ratio of two unstable nuclei by substituting $T_\mathrm{Gal}$ with the mean life of the reference isotope. However, our recommended factors for Equation~\ref{eq:steady} are not applicable in this case because they are based on GCE calculations of an unstable-to-stable ratio. For $^{247}$Cm/$^{235}$U, using the GCE code results in shorter isolation times by 22\,\%, 12\,\%, and 8\,\% for the maximum, best, and minimum predictions, respectively.

From Table~\ref{tab:slr2}, the results from the radionuclides produced exclusively by the $r$ process ($^{129}$I and $^{247}$Cm) confirm the previous results of isolation times consistent with each other, in particular when considering the maximum prediction, ranging from 86 to roughly 120\,Myr. When considering the other $r$-process short-lived radionuclide $^{244}$Pu (with half life 80\,Myr), as in the case of $^{247}$Cm, Equation~\ref{eq:steady} is not valid because the reference isotope in this case is the unstable $^{238}$U, with a mean life of roughly 6.5 Gyr, and the isolation times are always longer when calculated using the code. Results on the isolation times derived using this isotope are broadly consistent with those of the other two $r$-process isotopes, however, they are not reported in the table because the ESS ratio in this case is not determined well enough yet to be able to give accurate values. The results for the radionuclides produced also by the $s$ process ($^{107}$Pd and $^{182}$Hf) give isolation times consistent with each other, between 27 and 44\,Myr, but much shorter than those derived from the $r$-process nuclei \citep{lugaro14}.

This discrepancy indicates the limitation of assuming a continuous stellar production rate in the Galaxy, which cannot accurately represent the small-scale temporal (order of tens of Myr) and spatial (order of a few parsec at most) inhomogeneities in the ISM related to the formation of the Sun. In our framework of continuous enrichment and homogeneous ISM, the material from which the Sun formed was apparently isolated from different nucleosynthetic sources at different times. This is because, as discussed in Sec.~\ref{sec:disc_future}, in reality these sources contributed in a discrete way, each with a different typical recurrent timescale $\delta$. Such recurrent timescale must be by definition longer than the isolation times calculated here: i.e., the $\delta$ related to the $r$ and the $s$ process should be longer than $\sim$\,80\,Myr and $\sim$\,30\,Myr, respectively. This difference agrees qualitatively with the fact that the $r$-process sources in the Galaxy (neutron star mergers and special supernovae) are expected to be less common than the $s$-process source (AGB stars of initial mass in the range roughly 2 to 4\,$M_{\odot}$). This topic needs to be further investigated using statistical means, as well as more sophisticated codes.

In relation to the $p$-process nuclei shown in Table~\ref{tab:slr1} ($^{92}$Nb and $^{146}$Sm), the picture is much less clear. The first problem is that the half life of $^{146}$Sm is uncertain, and if we use the two different currently proposed values we obtain very different results. The half life is a crucial parameter because it affects the isolation time both linearly via the free decay law and logarithmically via the abundance ratio calculation. Furthermore, due to the relatively long half life of $^{146}$Sm, the GCE uncertainties result in much larger uncertainties in the isolation time, up to an order of magnitude if we use Equation~\ref{eq:steady}. We also note that for this isotope the differences between the simple equation and our GCE code are very large, up to a factor of 6 in the abundance ratios, which is another effect of the relatively long half life. Furthermore, the potential origin(s) of the $p$-process nuclei in the Galaxy is still very uncertain, with both core-collapse and Type Ia supernovae being proposed. In the table, we considered Type Ia supernovae as the source of both isotopes, but this is unlikely \citep{travaglio18} and it leads to completely inconsistent isolation times. If we consider contributions of half of the $^{92}$Nb and $^{92}$Mo in the Galaxy from Type Ia supernovae and half from core-collapse supernovae and use a production ratio of 0.0082 for the latter \citep[from][]{lugaro16}, we obtain isolation times roughly between 20 and 80\,Myr. However, this is a purely speculative test. Due to all these issues, we cannot at the moment make any strong conclusion on the source of the $p$-process short-lived radioactive nuclei in the ESS and the derived isolation times.

Finally, we consider the shortest lived isotopes in Table~\ref{tab:slr1}: $^{26}$Al, $^{53}$Mn, and $^{60}$Fe. We confirm all previous conclusions that the ESS abundance of $^{26}$Al cannot be explained by the chemical evolution of the Galaxy. This conclusion holds even if we multiply the production ratio by a factor of ten. On the other hand, the abundances of both $^{53}$Mn and $^{60}$Fe could have been inherited from the ISM, leading to isolation times of the order to 10 to 20\,Myr from the supernova processes that produced them. In the code, we considered delay times corresponding to the single-degenerate scenario for Type Ia supernova. However, we tested the potential effect of delay times corresponding to the double-degenerate scenario for Type Ia supernovae for $^{53}$Mn and $^{60}$Fe and found a slight increase in the isolation times with respect to using the single-degenerate scenario. The difference in the isolation times derived from the two different isotopes could potentially be ascribed to them having different main Galactic sources: Type Ia supernovae for $^{53}$Mn and core-collapse supernovae for $^{60}$Fe. However, since the contribution and the yields of the two different supernova sources are still uncertain, we do not draw major conclusions here on the potential isolation time related to supernovae.

\section{Summary and Conclusion}
\label{sec:conc}
We presented an extension of the open-source GCE codes \texttt{SYGMA} (\citealt{2017arXiv171109172R}), \texttt{OMEGA} (\citealt{2017ApJ...835..128C}), and \texttt{OMEGA+} (\citealt{2017arXiv171006442C}), which allows to follow the decay of radioactive isotopes in the ISM. Our codes are connected to a decay module that includes 22 different decay channels and keeps track of any radioactive isotope of interest for GCE. Our framework can be used to predict the average isotopic composition of the ISM at the time the Sun formed, a key requirement in studying the origin of our Solar System and interpreting the presence of radioactive isotopes in the early Solar System, as inferred by meteorite data analysis.

In this paper we focused on the general evolution of isotopic mass ratios ($M_\mathrm{radio}/M_\mathrm{stable}$) that involve a radioactive and a stable isotope. We described in detail how the predicted evolution of such ratios in the Milky Way depends on the assumptions made for the SFH, the amount of gas present in the Galactic disk, the delay-time distribution of the nucleosynthesis sources, and the strength of galactic outflows. By the time the Sun formed, our predictions for radioactive-to-stable isotope ratios are uncertain by a factor of 3.6, given the uncertainties in the observations used to calibrate our Milky Way model. 
The evolution of isotopic ratios involving two radioactive isotopes on the other hand are less uncertain. For example, in the case of $^{235}$U\,/\,$^{238}$U our prediction by the time the Sun formed is uncertain by a factor of 60\,\%, and in the case of $^{60}$Fe\,/\,$^{26}$Al our prediction are almost devoid of GCE uncertainty. Ratios involving two short-lived radioactive isotopes thus offer the best conditions to probe and constrain nuclear astrophysics and the nucleosynthesis of radioactive isotopes, at least within a continuous and homogenized enrichment scenario. But for isotopic ratios involving a stable isotope ($M_\mathrm{radio}/M_\mathrm{stable}$), galaxy evolution and nuclear astrophysics uncertainties (not considered here) can affect the ratios in a similar way.

The result of our best-fit model for the  $M_\mathrm{radio}/M_\mathrm{stable}$ ratio by time of the formation of the Sun is similar to the result obtained by steady-state equation (Equation~\ref{eq:steady}), but multiplied by a factor of $2.3^{+3.4}_{-0.7}$. However, to account for the impact of metallicity- and mass-dependent yields, our numerical framework must be used instead of the steady-state equation. This capability, which will be addressed in future studies, aims to reinforce the connection between the fields of nuclear astrophysics, cosmochemistry, and meteorite data analysis. 

The tools presented in this work provide an ideal framework for future studies, including the statistical investigation of all the uncertainties, from the nuclear input for the decay rates, to the stellar yields, to the GCE observational constraints. Our codes will also allow to investigate all possible radioactive isotopes of interest simultaneously, from those with half-lives in the range of $0.1-1$\,Myr all the way to uranium isotopes with half-lives of the order of Gyr. As a preliminary test, we have calculated the isolation time of Solar System matter from the ISM on the basis of several radioisotopes well known to be present in the early Solar System. We confirm the dichotomy between nuclei with an $r$-process origin only and nuclei with both an $r$- and $s$-process origin. In relation to the $p$-process nuclei, too many uncertainties prevent us from drawing any preliminary conclusions. We also confirm the fact that $^{26}$Al in the early Solar System cannot be explained by Galactic chemical evolution, while $^{55}$Mn and $^{60}$Fe can.

\acknowledgments

We are grateful to the anonymous referee for her/his request to include in the paper Tables~\ref{tab:slr2} and \ref{tab:slr1} and Section~\ref{sec:disc_ESS}. This research is supported by the ERC Consolidator Grant (Hungary) funding scheme (Project RADIOSTAR, G.A. n. 724560). BKG and MP acknowledge the support of STFC, through the University of Hull Consolidated Grant ST/R000840/1, and access to {\sc viper}, the University of Hull High Performance Computing Facility. BC also acknowledges the support from the National Science Foundation (NSF,
USA) under grant No. PHY-1430152 (JINA Center for the Evolution of the
Elements). This research has received funding from the European Research Council under the European Union's Seventh Framework Programme (FP/2007-2013) / ERC Grant Agreement n. 615126.

\software{
\texttt{SYGMA} \citep{2017arXiv171109172R},
\texttt{OMEGA+} \citep{2017arXiv171006442C},
\texttt{NumPy} \citep{2011arXiv1102.1523V},
\texttt{matplotlib} (\url{https://matplotlib.org}).}

%

\vspace{5mm}


\appendix
\section{Parameters in Clayton's Model}
\label{ap_1}
In the analytical model of \cite{clayton84, clayton88}, the galactic inflow rate is defined by

\begin{equation}
    \dot{M}_\mathrm{inflow}(t) = \frac{k}{t+\Delta}M_\mathrm{gas}(t),
\end{equation}
where $k$ and $\Delta$ are free parameters. The star formation rate is given by 

\begin{equation}
    \dot{M}_\star(t) = \frac{\omega}{1-R}M_\mathrm{gas}(t),
\end{equation}
where $\omega$ and $R$ represent the gas consumption rate and the fraction of stellar mass returned into the ISM by dying stars. As shown in Section~\ref{sec:SFH}, the shape of the SFH plays an important role on the evolution of the $M_\mathrm{radio}/M_\mathrm{stable}$ ratio. In that section, we ran three models with Clayton's inflow prescription using $k=0$, 1, and 2, and assuming $\Delta=0.5$\,Gyr. We tuned the initial mass of gas and the parameter $\omega$ to ensure that all three models form the same amount of stars and end up with the same amount of gas. With this setup, using larger $k$ values pushes the peak of star formation to later times (Figure~\ref{fig:SFH}). However, this is not a general statement, as the $\Delta$ parameter can also change the shape of the SFH. 

\begin{figure}
\center
\includegraphics[width=3.35in]{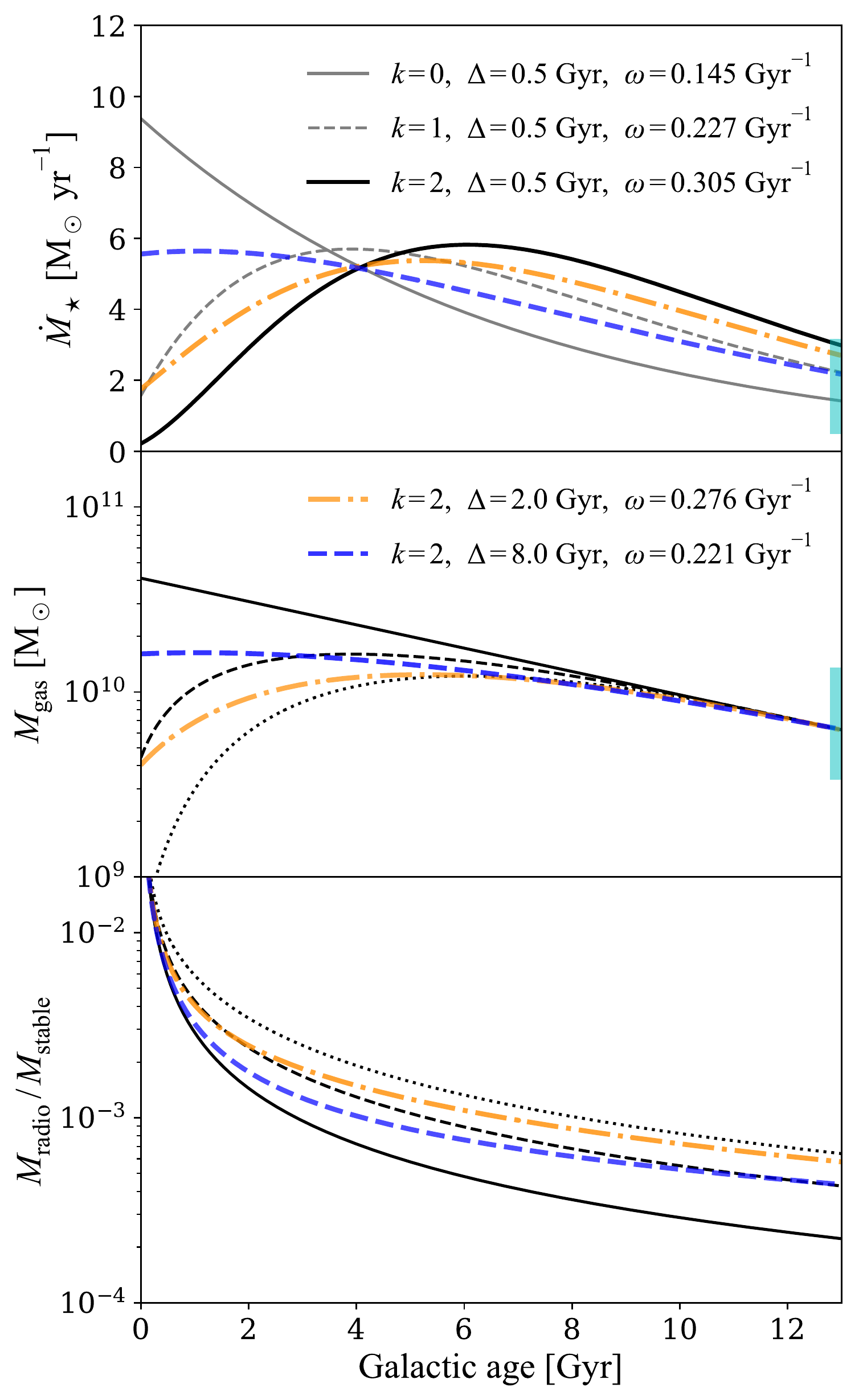}
\caption{Evolution of the star formation rate (top panel), mass of gas (middle panel), and isotopic mass ratio (lower panel), generated by integrating the system of equations of \cite{clayton84,clayton88} and by using different input parameters as labelled in the top and middle panels. We refer to Appendix~\ref{ap_1} for a description of the parameters. The cyan bands at 13~Gyr are observational constraints taken from \cite{2015A&A...580A.126K}.
\label{fig:clayton}}
\end{figure}

Figure~\ref{fig:clayton} shows the results of three models with $\Delta=0.5$\,Gyr and different $k$ values. Those are similar to the black lines shown in Figure~\ref{fig:SFH}, but here they are entirely computed using Clayton's equations, they are not generated by \texttt{OMEGA+}. We also added in Figure~\ref{fig:clayton} two additional models with $k=2$ and $\Delta=2$ and 8\,Gyr, with tuned values for $\omega$. When keeping $k$ constant, using a larger $\Delta$ pushes back the peak of star formation to earlier times, which means that even with the same $k$ (here $k=2$), it is possible to create variations in the $M_\mathrm{radio}/M_\mathrm{stable}$ ratio (lower panel of Figure~\ref{fig:clayton}). This statement may appear to be in contradiction with Clayton's widely used analytical approximation,

\begin{equation}
    \frac{M_\mathrm{radio}}{M_\mathrm{stable}}(t)=(k+1)\frac{P_\mathrm{radio}}{P_\mathrm{stable}}\frac{\tau}{t},
\label{eq_clayton}
\end{equation}
in which $k$ is the only galaxy evolution parameter that can alter the isotopic ratio. But this analytical solution is only valid when $\Delta \ll t$ (see \citealt{huss09} for mathematical development) and cannot be applied when $\Delta$ is of the order of a few Gyrs. The results shown in Figure~\ref{fig:clayton} and the orange lines in Figure~\ref{fig:SFH} were all generated by integrating Clayton's system of equations, we did not use the approximated solution.

All models shown in this section have the same final gas-to-star mass ratio. For a given $k$, different combinations of $\Delta$ and $\omega$ parameters can thus recover the same observational constraint. This degeneracy has also been highlighted by \citet{huss09} for $\Delta=0.1$ and 0.5\,Gyr (see their Table~2).



\bibliographystyle{yahapj}
\bibliography{ms}



\end{document}